\documentclass[preprint,prd,aps,showpacs,showkeys,nofootinbib]{revtex4}
\usepackage{graphicx}
\usepackage{amsmath,amssymb,amsfonts}
\usepackage{float}
\usepackage{color}
\begin{document}


\title{Pair production of neutral Higgs particles in the B-LSSM }

\author{Dan He$^{a,b}$\footnote{hedandeya@163.com},
Tai-Fu Feng$^{a,b,d}$\footnote{fengtf@hbu.edu.cn},
Jin-Lei Yang$^{c,e}$\footnote{yangjinlei@itp.ac.cn},
Guo-Zhu Ning$^{a, b}$,
Hai-Bin Zhang$^{a,b}$\footnote{hbzhang@hbu.edu.cn},
Xing-Xing Dong$^{a, b}$}

\affiliation{$^a$Department of Physics, Hebei University, Baoding, 071002, China\\
$^b$Key Laboratory of High-precision Computation and Application of Quantum Field Theory of Hebei Province, Baoding, 071002, China\\
$^c$CAS Key Laboratory, Institute of Theoretical Physics, Chinese Academy of Sciences, Beijing, 100190, China\\
$^d$College of Physics, Chongqing University, Chongqing, 400044, China\\
$^e$School of Physical Sciences, University of Chinese Academy of Sciences, Beijing, 100049, China}

\begin{abstract}
Higgs pair production provides a unique handle for measuring the strength of Higgs
self interaction and constraining the shape of the Higgs potential.
Including radiative corrections to the trilinear couplings of $CP$-even Higgs,
we investigate the  cross section of the lightest neutral Higgs pair production in gluon
fusion at the Large Hadron Collider in the supersymmetric extensions of the standard model.
Numerical results indicate that the correction to the cross section is about 11\% in the B-LSSM,
while is only about 4\% in the MSSM.
Considering the constraints of the experimental data of the lightest Higgs,
we find that the gauge couplings of $U(1)_{B-L}$ and the ratio of the nonzero
vacuum expectation values of two singlets also affect strongly the theoretical evaluations on
the production cross section in the B-LSSM.
\end{abstract}


\maketitle

\section{Introduction\label{sec1}}
The discovery of the Higgs boson at the Large Hadron Collider (LHC)~\cite{mh-ATLAS,mh-CMS} is a great triumph
of the standard model (SM). Combining updated data~\cite{mh-LHC,mh-CMS1,mh-ATLAS1}, one obtains the measured
mass of the Higgs as~\cite{PDG2}
\begin{eqnarray}
&&m_h=125.25\pm 0.17\: {\rm{GeV}}.
\label{mh-exp}
\end{eqnarray}
The precise electroweak observable of Higgs boson set some stringent constraints on parameter space of
the SM and its various extensions. In addition, the SM cannot provide natural explanations of some problems,
such as the neutrino masses, the hierarchy problem, the Dark Matter(DM) candidates etc.
The aforementioned problems can be accommodated in some new physics extensions of the SM, in which
the various supersymmetric extensions of the SM are popular. However, the minimal supersymmetric extension of
the SM (MSSM) cannot explain the tiny masses of neutrino naturally.
To obtain the lightest $CP$-even Higgs with mass around $125\;{\rm{GeV}}$ in the
MSSM, we should take into account the large radiative corrections
from the third generation quark and their super partners because the tree level mass of the lightest
$CP$-even Higgs is less than $m_{_{\rm Z}}$. Meanwhile, masses of other neutral scalar
and charged Higgs are much heavier than the mass of lightest Higgs, namely $125\;{\rm{GeV}}$. This is known as
the little hierarchy problem in the MSSM~\cite{hierarchy problem}.
Gauge group of the supersymmetry extension of the SM with local $B-L$ symmetry (B-LSSM)~\cite{BLSSM,BLSSM1}
is $U(1)_Y \bigotimes SU(2)_L \bigotimes SU(3)_C\bigotimes U(1)_{B-L}$,
where $B$ and $L$ stand for the baryon number and the lepton number, respectively.
The B-LSSM provide an elegant explanation for the tiny masses of the left-handed neutrino through seesaw mechanism,
and the B-LSSM also alleviates the little hierarchy problem of the MSSM,  because the exotic singlet
Higgs and right-handed (s)neutrinos~\cite{sn,sn1,sn2,sn3,sn4,sn5,sn6,sn7} release additional
parameter space from the LEP, Tevatron and LHC constraints.
Moreover, the model can also provide more DM candidates than that of the MSSM~\cite{DM,DM1,DM2,DM3}.

Higgs pair production provides a unique handle for measuring the strength
of the Higgs self interaction and constraining the shape of the Higgs potential.
In the SM, the Higgs potential is
\begin{eqnarray}
V=-\mu^2\phi^\dagger\phi+\lambda(\phi^\dagger\phi)^2\;,
\label{Higgs boson potential}
\end{eqnarray}
which is completely specified by two parameters $\mu$ and $\lambda$. $\mu$ and $\lambda$ can be determined from
the vacuum expectation value (VEV) of the Higgs field, and the mass of the Higgs boson,
but there is no direct measurement beyond that.
The next step in understanding the
shape of the Higgs potential is to measure the Higgs trilinear coupling
which can be probed by the neutral Higgs boson pair production at the LHC.
In scenarios of the SM the Higgs pair production at the LHC proceeds by the parton process $gg\rightarrow hh$
through the heavy quark induced box diagrams and also through the production of an off-shell Higgs
which subsequently splits into two on-shell Higgs~\cite{SMP1,SMP2,SMP3}. If radiative corrections of top quark to
the amplitude of the parton process $gg\rightarrow hh$ from box diagrams are determined through
other experimental data, the production cross-section depends on Higgs trilinear self-coupling sensitively
~\cite{Higgs-coupling1,Higgs-coupling2,Higgs-coupling3,Higgs-coupling4,Higgs-coupling5,
Higgs-coupling6,Higgs-coupling7,Higgs-coupling8,Higgs-coupling9}.
At the center of mass energy of $14\:{\rm{TeV}}$, the theoretical evaluation of the production
cross section is about $17\:{\rm{fb}}$ at the leading order (LO),
and reaches roughly $35\:{\rm{fb}}$ after including the next-to-leading order (NLO) QCD
correction~\cite{QCD-correction,QCD-correction1} in the SM.
In the MSSM the radiative corrections of bottom quark to the amplitude of the parton process $gg\rightarrow hh$ of
box diagrams may be enhanced by the ratio $\tan\beta=\upsilon_{_2}/\upsilon_{_1}$ between the nonzero
VEVs of Higgs fields~\cite{quark-mssm,quark-mssm1,quark-mssm2}.

In the B-LSSM, the tree level prediction for the mass of the lightest $CP$-even Higgs
can exceed $m_{_{\rm Z}}$ since the spectrum includes four $CP$-even Higgs.
Based on the present experimental data, the lower bound on the mass of the next-to-light neutral Higgs can be set as
$135\;{\rm GeV}$~\cite{next-to-lightest Higgs,next-to-lightest Higgs1,next-to-lightest Higgs2}.
Thus, the pair production of the lightest Higgs receives additional contributions
from the parton process $gg\rightarrow h_i\rightarrow hh$ with $h_i$ denoting a $CP$-even non-standard Higgs
~\cite{SM-higgs,SM-higgs1,SM-higgs2,SM-higgs3,SM-higgs4,SM-higgs5,SM-higgs6,
SM-higgs7,SM-higgs8,SM-higgs9,SM-higgs10,SM-higgs11,SM-higgs12,SM-higgs13,SM-higgs14}. Furthermore, the trilinear-coupling
of the lightest Higgs is also modified drastically from the mixing between the doublets and singlets together
with radiative corrections to the scalar potential.

The paper is organized as follows. In Sec~\ref{sec2}, we briefly present the features of the B-LSSM
and calculate the radiative corrections to the trilinear couplings of $CP$-even Higgs.
In Sec~\ref{sec3}, we calculate pair production cross section of the lightest Higgs
at the NLO approximation. The numerical analysis is performed in Sec~\ref{sec4},
and conclusions are summarized in Sec~\ref{sec5}. The tedious formulae are collected in Appendices.

\section{the B-LSSM\label{sec2}}
In this section, we briefly introduce the basic properties of the B-LSSM, and then present the radiative
corrections to the pole masses and trilinear couplings of $CP$-even Higgs, respectively.
Finally we present a concise discussion of the possible constraints on model parameters
imposed by $\Bar{B}\rightarrow X_{_s}\gamma$, $B_{_s}^0\rightarrow\mu^+\mu^-$ and LHC data.
Here we adopt the concrete model described in Refs.~\cite{BLSSMB,BLSSMB1,BLSSMB2,BLSSMB3} to proceed with our analysis,
where the chiral superfields and their quantum numbers are listed in Table.~\ref{tab1}.
In order to break down the local symmetry $U(1)_{B-L}$ spontaneously, two chiral singlet superfields $\hat\eta_1$, $\hat\eta_2$
are introduced. In addition, the couplings among three generations of right-handed neutrinos
and the singlet $\hat\eta_1$ provide the see-saw mechanism to produce the tiny masses of left-handed
neutrinos.

The corresponding superpotential of the B-LSSM is written as
\begin{eqnarray}
W=W_{MSSM}+Y_{\nu,_{ij} }\hat{L_i}\hat{H_2}{\hat\nu_j}^c-\mu^\prime{\hat\eta}_1\hat{\eta}_2
+Y_{x,ij}{\hat\nu_i}^c\hat\eta_1 {\hat\nu_j}^c\;,
\label{superpotential}
\end{eqnarray}
where $W_{MSSM}$ is the superpotential of the MSSM, and $i,\;j$ are generation indices.
To break the supersymmetry, the soft breaking terms are generally given by
\begin{eqnarray}
&&\mathcal{L}_{soft}=\mathcal{L}_{MSSM}+[-M_{BB'}\tilde{\lambda}_{B'}\tilde{\lambda}_B
-\frac{1}{2}M_{B'}\tilde{\lambda}_{B'}\tilde{\lambda}_{B'}
-B_{\mu'}\tilde{\eta}_1\tilde{\eta}_2
+T_{\nu}^{ij} H_2\tilde{\nu}_i^c\tilde{L}_j
\nonumber\\&&\hspace{1.2cm}+T_x^{ij}\tilde{\eta}_1\tilde{\nu}_i^c\tilde{\nu}_j^c+h.c]
-m_{\tilde{\nu},ij}^2(\tilde{\nu}_i^c)^*\tilde{\nu}_j^c
-m_{\tilde{\eta}_1}^2|\tilde{\eta}_1|^2
-m_{\tilde{\eta}_2}^2|\tilde{\eta}_2|^2\;,
\label{soft}
\end{eqnarray}
where $\mathcal{L}_{MSSM}$ is the soft breaking terms of the MSSM, and $\tilde{\lambda}_{B}$, $\tilde{\lambda}_{B'}$
represent the gauginos of $U(1)_Y$ , $U(1)_{B-L}$ correspondingly.
\begin{table*}
\begin{tabular*}{\textwidth}{@{\extracolsep{\fill}}lllll@{}}
\hline
superfields&Spin0&Spin{1/2}&$U(1)_Y \bigotimes SU(2)_L \bigotimes SU(3)_C\bigotimes U(1)_{B-L}$\\
\hline
$\hat Q$&$\tilde{Q}$&Q&\qquad\qquad\qquad($\frac{1}{6}$,2,3,$\frac{1}{6}$)\\
$\hat D$&{$\tilde{d}^c$}&$d^c$&\qquad\qquad\qquad($\frac{1}{3}$,1,$\bar{3}$,$-{\frac{1}{6}}$)\\
$\hat U$&{$\tilde{u}^c$}&$u^c$&\qquad\qquad\qquad($-{\frac{2}{3}}$,1,$\bar{3}$,$-{\frac{1}{6}}$)\\
$\hat L$&{$\tilde{L}$}&$L$&\qquad\qquad\qquad($-{\frac{1}{2}}$,2,1,$-{\frac{1}{2}}$)\\
$\hat E$&{$\tilde{e}^c$}&$e^c$&\qquad\qquad\qquad(1,1,1,${\frac{1}{2}}$)\\
${\hat\nu}$&{$\tilde{\nu}^c$}&${\nu}^c$&\qquad\qquad\qquad(0,1,1,${\frac{1}{2}}$)\\
${\hat H}_1$&{$H_1$}&{$\tilde{H}_1$}&\qquad\qquad\qquad($-{\frac{1}{2}}$,2,1,0)\\
${\hat H}_2$&{$H_2$}&{$\tilde{H}_2$}&\qquad\qquad\qquad(${\frac{1}{2}}$,2,1,0)\\
${\hat{\eta}}_1$&{$\eta_1$}&{$\tilde{\eta}_1$}&\qquad\qquad\qquad(0,1,1,-1)\\
${\hat{\eta}}_2$&{$\eta_2$}&{$\tilde{\eta}_2$}&\qquad\qquad\qquad(0,1,1,1)\\
\hline
\end{tabular*}
\caption{Chiral superfields and their quantum numbers in the B-LSSM.}
\label{tab1}
\end{table*}

The local gauge symmetry $SU(2)_L \bigotimes U(1)_Y \bigotimes U(1)_{B-L}$ is broken down to the electromagnetic
symmetry $U(1)_{em}$ when the Higgs fields receive nonzero VEVs:
\begin{eqnarray}
&&H_1^1=\frac{1}{\sqrt{2}}(v_1+{\phi_d}+\emph{i}\rm{Im}{H_1^1}),\nonumber\\
&&H_2^2=\frac{1}{\sqrt{2}}(v_2+\phi_u+\emph{i}\rm{Im}{H_2^2}),\nonumber\\
&&\tilde{\eta}_1=\frac{1}{\sqrt{2}}(u_1+\phi_{\tilde{\eta}_1}+\emph{i}\rm{Im}{\tilde\eta_1}),\nonumber\\&&
\tilde{\eta}_2=\frac{1}{\sqrt{2}}(u_2+\phi_{\tilde{\eta}_2}+\emph{i}\rm{Im}{\tilde\eta_2})\;.
\label{VEV}
\end{eqnarray}
where the mixing among $\phi_d,\;\phi_u,\;\phi_{\tilde{\eta}_1},\;\phi_{\tilde{\eta}_2}$ produces
four $CP$-even neutral eigenstates of Higgs, and the mixing among $\rm{Im}{H_1^1},\;\rm{Im}{H_2^2},\;
\rm{Im}{\tilde\eta_1},\;\rm{Im}{\tilde\eta_2}$ produces two neutral Goldstones and two $CP$-odd neutral
eigenstates of Higgs, respectively.
For convenience, we define $u^2={u^2_1}+{u^2_2}$, $v^2={v^2_1}+{v^2_2}$, and ${\tan{\beta^\prime}}=u_2/u_1$
in analogy to the ratio of two nonzero VEVs of doublets.

The effective potential can be written as~\cite{Veff}
\begin{eqnarray}
V_{eff}=V_0+\Delta{V}\;,
\label{eff-potential}
\end{eqnarray}
where $V_0$ denotes the scalar potential at the tree level, and $\Delta{V}$ denotes the one-loop radiative
corrections to the effective potential. The concrete expression of the tree level scalar potential is
\begin{eqnarray}
&&V_0=(m^2_{H_1}+\mu^2){|H_1^1|}^2+(m^2_{H_2}+\mu^2){|H_2^2|}^2+(m^2_{\tilde\eta_1}+\mu'^2)|{\tilde\eta_1}|^2+(m^2_{\tilde\eta_2}+\mu'^2)|{\tilde\eta_2}|^2\nonumber\\
&&\quad\:\:\:\;\;-2{B_\mu} H_1^1 H_2^2-2{B_{\mu'}} {\tilde\eta_1}{\tilde\eta_2} +\frac{1}{8} g^2 ({|H_1^1|}^2-{|H_2^2|}^2)^2+\frac{1}{2}{g^2_{_B}}{(|{\tilde\eta_1}|^2-|{\tilde\eta_2}|^2)^2}\nonumber\\
&&\quad\:\:\:\;\;+\frac{1}{2} g_{_B} g_{_{YB}}({|H_1^1|}^2-{|H_2^2|}^2)(|{\tilde\eta_1}|^2-|{\tilde\eta_2}|^2)\;.
\label{Tree-potential}
\end{eqnarray}
In the above we have adopted the abbreviation $g^2=g^2_{_1}+g^2_{_2}+g^2_{_{YB}}$, where $g_{_2}$
denotes the gauge coupling of $SU(2)_L$, $g_{_1}$ denotes the gauge coupling of $U(1)_{_Y}$,
$g_{_B}$ denotes the gauge coupling of $U(1)_{_{B-L}}$, and $g_{_{YB}}$ denotes the mixing
coupling between $U(1)_{_Y}$ and $U(1)_{_{B-L}}$ in the covariant derivative, respectively.
The radiative corrections to the effective potential is dominated by
the contributions from top quark, scalar top quark, bottom quark and scalar bottom quark.
The concrete expression for the one-loop effective potential $\Delta{V}$ reads
\begin{eqnarray}
&&\Delta{V}=-\frac{3}{64\pi^2}{m_{_t}^4}[\ln(\frac{m_{_t}^2}{Q^2})-\frac{3}{2}]
-\frac{3}{64\pi^2}{m_{_b}^4}[\ln(\frac{m_{_b}^2}{Q^2})-\frac{3}{2}]
\nonumber\\
&&\hspace{1.2cm}
+\frac{3}{128\pi^2}\sum_{i=1}^{2}{m_{\tilde{t}_{i}}^4}[\ln(\frac{m_{\tilde{t}_{i}}^2}{Q^2})-\frac{3}{2}]
+\frac{3}{128\pi^2}\sum_{i=1}^{2}{m_{\tilde{b}_{i}}^4}[\ln(\frac{m_{\tilde{b}_{i}}^2}{Q^2})-\frac{3}{2}]\;.
\label{1Loop-potential}
\end{eqnarray}

In the basis (${\phi_d}$,${\phi_u}$,$\phi_{\tilde{\eta}_1}$,$\phi_{\tilde{\eta}_2}$) the tree level mass
squared matrix for $CP$-even neutral Higgs bosons is given by
\begin{eqnarray}
&&\Big[m_{_h}^2\Big]^{(0)}=\left(\begin{array}{cc}\Big[m_{_h}^2\Big]_{_{\phi\phi}},\;\;&\Big[m_{_h}^2\Big]_{_{\phi\phi_{_{\tilde\eta}}}}\\
\Big[m_{_h}^2\Big]_{_{\phi\phi_{_{\tilde\eta}}}}^T,\;\;&\Big[m_{_h}^2\Big]_{_{\phi_{_{\tilde\eta}}\phi_{_{\tilde\eta}}}}
\end{array}\right)\;,
\label{mh-tree-level}
\end{eqnarray}
where the superscript $T$ denotes the transpose operation,
and the explicit form of the $2\times2$ submatrices are given in the appendix~\ref{app-A}.

As stating in Ref.~\cite{GDegrassi}, the nonzero VEVs and parameters in the tree level scalar potential
are $\overline{\rm MS}$ renormalized quantities. Neglecting width effects, one acquires the pole masses
of neutral $CP$-even Higgs through the pole equation
\begin{eqnarray}
&&{\rm Det}\Big[m_{_h}^2-\frac{{\partial^2V_{_{eff}}}}{{\partial\phi_{_i}\partial\phi_{_j}}}\Big|_{\phi_{_i}=\upsilon_{_i}}
-\Pi_{_{\phi_{_i}\phi_{_j}}}(m_{_h}^2)\Big]=0\;,
\label{pole-equation}
\end{eqnarray}
where $\Pi_{_{\phi_{_i}\phi_{_j}}}(p^2)$ is the self-energy (two-point) functions of the
$CP$-even Higgs with external 4-momentum $p$.
Taking the approach adopted in Ref.\cite{GDegrassi}, we rewrite the radiative corrections to
the mass squared matrix of CP-even Higgs as
\begin{eqnarray}
&&\frac{{\partial^2V_{_{eff}}}}{{\partial\phi_{_i}\partial\phi_{_j}}}\Big|_{\phi_{_i}=\upsilon_{_i}}
+\Pi_{_{\phi_{_i}\phi_{_j}}}(m_{_h}^2)
\nonumber\\
&&\hspace{-.6cm}
=\Big[\frac{{\partial^2V_{_{eff}}}}{{\partial\phi_{_i}\partial\phi_{_j}}}\Big|_{\phi_{_i}=\upsilon_{_i}}
+\Pi_{_{\phi_{_i}\phi_{_j}}}(0)\Big]+\Big[\Pi_{_{\phi_{_i}\phi_{_j}}}(m_{_h}^2)
-\Pi_{_{\phi_{_i}\phi_{_j}}}(0)\Big]
\nonumber\\
&&\hspace{-.6cm}
=\Big[m_{_h}^2\Big]_{_{\phi_{_i}\phi_{_j}}}^V+\Delta\Pi_{_{\phi_{_i}\phi_{_j}}}(m_{_h}^2)\;,
\label{pole-equation1}
\end{eqnarray}
where the dominant first term can be calculated in term of derivatives of the effective potential
$V_{_{eff}}$, and the second term is ultraviolet finite, which only induces contributions suppressed
by the small couplings and loop factor.


\begin{figure}
\setlength{\unitlength}{1mm}
\centering
\begin{minipage}[c]{0.9\textwidth}
\includegraphics[width=5.3in]{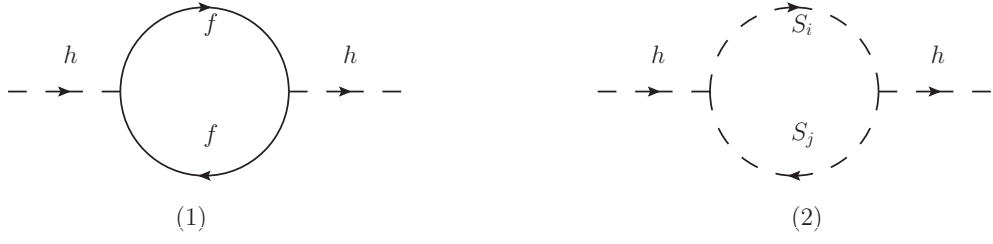}
\end{minipage}
\caption[]{One-loop diagrams which induce the radiative corrections to
the lightest Higgs boson mass, where $f=t,\;b$, and
$S_i=\tilde{t}_{i},\;\tilde{b}_{i}$ with $i=1,\;2$.}
\label{Higgs1}
\end{figure}
In order to get the pole mass of the lightest Higgs consistently, we write
the explicit expressions for the self energy diagrams. It is well known that
the dominant corrections to the lightest Higgs mass originate from bottom, sbottom, top and stop.
The relevant diagrams are plotted in Fig.\ref{Higgs1}, and the corresponding corrections are given by Eq.(\ref{self-corrections}).
Choosing the relevant parameters appropriately in the B-LSSM, we find that the particle spectrum
contains not only a neutral Higgs with a mass around $125\;{\rm{GeV}}$, but also a next-to-lightest
neutral Higgs with a mass of several hundreds of GeV whose main component originates from two singlets.

Including the radiative corrections, the trilinear
couplings among $CP$-even Higgs can be written similarly as
\begin{eqnarray}
&&{C_{h_ah_bh_c}}=\sum\limits_{\alpha,\beta,\gamma}\Big[S_{_{\alpha\beta\gamma}}
\frac{{\partial^3V_{_{eff}}}}{{\partial\phi_{_\alpha}\partial\phi_{_\beta}\partial\phi_{_\gamma}}}
[(Z_{_H})_{a\alpha}(Z_{_H})_{b\beta}(Z_{_H})_{c\gamma}
+(Z_{_H})_{a\alpha}(Z_{_H})_{c\beta}(Z_{_H})_{b\gamma}\nonumber\\
&&\hspace{0.8cm}+(Z_{_H})_{b\alpha}(Z_{_H})_{a\beta}(Z_{_H})_{c\gamma}+(Z_{_H})_{c\alpha}(Z_{_H})_{a\beta}(Z_{_H})_{b\gamma}
+(Z_{_H})_{b\alpha}(Z_{_H})_{c\beta}(Z_{_H})_{a\gamma}\nonumber\\
&&\hspace{0.8cm}+(Z_{_H})_{c\alpha}(Z_{_H})_{a\beta}(Z_{_H})_{b\gamma}+(Z_{_H})_{c\alpha}(Z_{_H})_{b\beta}(Z_{_H})_{a\gamma}]\Big]
+\Delta\Lambda_{_{abc}}(p_{_a}^2,\;p_{_b}^2,\;p_{_c}^2)\;,
\label{trilinear-couplings}
\end{eqnarray}
where $p$ denotes the external 4-momentum of the $CP$-even Higgs,
and the $4\times4$ matrix $Z_{_{H}}$ is the mixing between $\phi_{_d}$,
$\phi_{_u}$,$\phi_{\tilde{\eta}_1}$ and $\phi_{\tilde{\eta}_2}$.
In addition, the symmetry factor $S_{_{\alpha\beta\gamma}}$ is
\begin{eqnarray}
&&S_{_{\alpha\beta\gamma}}=\left\{\begin{array}{l}
1,\;\;\alpha\neq\beta\neq\gamma,\\
\frac{1}{2},\;\alpha=\beta\neq\gamma,\;\alpha=\gamma\neq\beta,\;\alpha\neq\beta=\gamma,\;\\
\frac{1}{6},\;\alpha=\beta=\gamma.\end{array}\right.
\label{symmetry-factor}
\end{eqnarray}

\begin{figure}
\setlength{\unitlength}{1mm}
\centering
\begin{minipage}[c]{0.9\textwidth}
\includegraphics[width=5.8in]{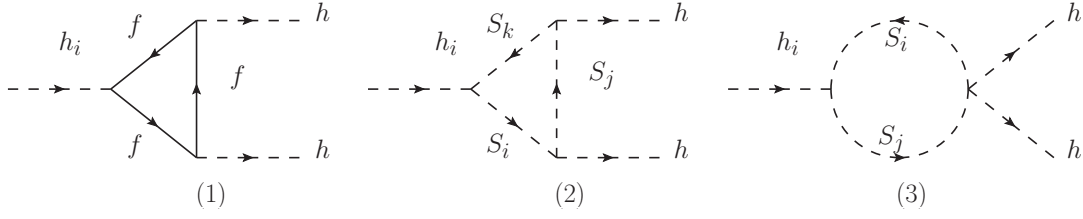}
\end{minipage}
\caption[]{The Feynman diagrams which induce the radiative corrections to
the trilinear couplings of CP-even Higgs, where $f=t,\;b$, and
$S_i=\tilde{t}_{i},\;\tilde{b}_{i}$ with $i=1,\;2$.}
\label{Higgs2}
\end{figure}

The second term in Eq.(\ref{trilinear-couplings}) is obtained through evaluation of
the triangle diagrams among the $CP$-even Higgs.
Feynman diagrams of dominant corrections to the trilinear couplings among
$CP$-even Higgs are plotted in Fig.\ref{Higgs2}, and
the concrete expressions of $\Delta\Lambda_{_{h_{_i}hh}}$ are collected in the appendix~\ref{app-B}.

The updated average experimental data on the branching ratios of $\bar{B}\rightarrow X_{_s}\gamma$
and $B_{_s}^0\rightarrow\mu^+\mu^-$ are~\cite{BaBar,Belle}
\begin{eqnarray}
&&Br(\bar{B}\rightarrow X_{_s}\gamma)=(3.49\pm0.19)\times10^{-4}\;,
\nonumber\\
&&Br(B_{_s}^0\rightarrow\mu^+\mu^-)=(2.9^{+0.7}_{-0.6})\times10^{-9}\;,
\label{RareB-1}
\end{eqnarray}
which set stringent constraints on the parameter space of the new physics extensions of the SM.
Generally the theoretical evaluation of the branching ratio of $\bar{B}\rightarrow X_{_s}\gamma$
is given by
\begin{eqnarray}
&&Br(\bar{B}\rightarrow X_{_s}\gamma)=R(C_{7\gamma}(\mu_{_b})+N(E_{\gamma}))\;,
\label{RareB-2}
\end{eqnarray}
where the overall factor $R=2.47\times10^{-3}$, and the nonperturbative contribution
$N(E_{_\gamma})=(3.6\pm0.6)10^{-3}$. The Wilson coefficient $C_{7\gamma}(\mu_{_b})$
at hadronic scale is
\begin{eqnarray}
&&C_{7\gamma}(\mu_{_b})=C_{7\gamma,SM}(\mu_{_b})+C_{7\gamma,NP}(\mu_{_b})\;.
\label{RareB-3}
\end{eqnarray}
Choosing the hadron scale $\mu_{_b}=2.5\;{\rm GeV}$, one obtains the SM contribution as
$C_{7\gamma,SM}(\mu_{_b})\simeq-0.3689$, and the new physics correction to the Wilson
coefficient at hadronic scale is
\begin{eqnarray}
&&C_{7\gamma,NP}(\mu_{_b})\simeq0.5696C_{7\gamma,NP}(\mu_{_{EW}})+0.1107C_{8\gamma,NP}(\mu_{_{EW}})\;,
\label{RareB-4}
\end{eqnarray}
where $\mu_{_{EW}}$ denotes the electroweak scale. Assuming that the supersymmetry partners
of quarks are heavier, we formulate the dominate corrections to the Wilson coefficients
at electroweak scale in the limit of large $\tan\beta$ as
\begin{eqnarray}
&&C_{7\gamma,NP}(\mu_{_{EW}})=\frac{{m_{_b}m_{_s}t_{_\beta}^2}}{{12m_{_{W}}^2}}
\Big[-2I_{_1}+I_{_3}-2I_{_4}\Big](x_{_t},\;x_{_{H^\pm}})+\cdots\;,
\nonumber\\
&&C_{8\gamma,NP}(\mu_{_{EW}})=\frac{{m_{_b}m_{_s}t_{_\beta}^2}}{{12m_{_{W}}^2}}
\Big[-I_{_1}+2I_{_3}-I_{_4}\Big](x_{_t},\;x_{_{H^\pm}})+\cdots\;,
\label{RareB-5}
\end{eqnarray}
where the concrete expressions of the functions $I_{_1},\;I_{_{3}},\;I_{_4}$
can be found in literature~\cite{tan}.

In the limit of large $\tan\beta$ and heavy supersymmetric particles, the leading corrections
to the effective lagrangian is given by the Wilson coefficient of electroweak scale
$C_{_{S}}(\mu_{_{EW}})$
\begin{eqnarray}
&&C_{_{S}}(\mu_{_{EW}})=\frac{m_{_t}^2m_{_b}m_{_\mu}t_{_\beta}^2}{4m_{_{W}}^2m_{_{h_0}}^2s_{_{W}}^2}
\Big[\frac{2(Z_{_H})_{_{11}}C_{_{h_0H^+H^-}}s_{_W}}{ e}\frac{\partial}{\partial x_{_{H^\pm}}}\varrho_{_{1,1}}
\nonumber\\
&&\hspace{2.2cm}
+(Z_{_H})_{_{11}}(Z_{_H})_{_{21}}\Big(\frac{\partial}{\partial x_{_t}}\varrho_{_{2,1}}
+x_{_t}\frac{\partial}{\partial x_{_t}}\varrho_{_{1,1}}\Big)\Big](x_{_t},x_{_{H^\pm}})+\cdots
\label{RareB-7}
\end{eqnarray}
where the definition of $\varrho_{_{m,n}}(x,y)$ can also be found in literature~\cite{tan}.
Here $C_{_{h_0H^+H^-}}$ is the trilinear coupling between the lightest neutral Higgs
and charged Higgs pair, and its expression is presented in appendix~\ref{app-C}.
In the concrete analyses, we should also include the corrections of two-loop
Barr-Zee and rainbow type diagrams, respectively.

The experimental constraints on the $Z^\prime$ mass can be extracted from
the Drell-Yan production cross-section of fermion-antifermion pairs in the LHC.
In the narrow width approximation (NWA), the Drell-Yan production cross section
can be simplified as
\begin{eqnarray}
&&\sigma_{_{f\bar{f}}}\simeq \frac{1}{3}\sum\limits_{q=u,d}\Big(\frac{dL_{_{q\bar{q}}}}{ dm_{Z^\prime}^2}
\hat{\sigma}(q\bar{q}\rightarrow Z^\prime)\Big)Br(Z^\prime\rightarrow f\bar{f})
\label{zPrime-1}
\end{eqnarray}
where $dL_{_{q\bar{q}}}/dm_{Z^\prime}^2$ denotes the parton luminosity,
and $\hat{\sigma}(q\bar{q}\rightarrow Z^\prime)$ is the peak cross section
\begin{eqnarray}
&&\hat{\sigma}(q\bar{q}\rightarrow Z^\prime)=\frac{\pi}{12}g_{_B}^2\Big[\Big(g_{_V}^q\Big)^2
+\Big(g_{_A}^q\Big)^2\Big]\;.
\label{zPrime-2}
\end{eqnarray}
Couplings between massive $Z^\prime$ and the SM fermions are generally written as
\begin{eqnarray}
&&\frac{g_{_B}}{2}Z^\prime_\mu\bar{f}\gamma^\mu\Big(g_{_L}^fP_{_L}+g_{_R}^fP_{_R}\Big)f\;,
\label{zPrime-3}
\end{eqnarray}
where $g_{_{V,A}}^f=\pm g_{_L}^f+g_{_R}^f$.


Suppose in the final state, there are only SM fermions.
The NNLO Drell-Yan cross section can be written as
\begin{eqnarray}
&&\sigma_{l^+l^-}^{NNLO}\simeq K_{_{NNLO}}^{PDF}\sigma_{l^+l^-}^{LO}\;,
\label{zPrime-7}
\end{eqnarray}
where the QCD factor $K_{_{NNLO}}^{PDF}$ originates from the NNLO QCD corrections
to the PDF's, and the LO cross section is
\begin{eqnarray}
&&\sigma_{l^+l^-}^{LO}=\frac{\pi}{48s}\Big\{c_uw_u(s,m_{Z^\prime}^2)+c_dw_d(s,m_{Z^\prime}^2)\Big\}\;.
\label{zPrime-8}
\end{eqnarray}
The coefficients $c_{u,d}$ are defined as
\begin{eqnarray}
&&c_u=\frac{g_{_B}^2}{2}\Big[\Big(g_{_V}^u\Big)^2+\Big(g_{_A}^u\Big)^2\Big]Br(Z^\prime\rightarrow l^+l^-)\;,
\nonumber\\
&&c_d=\frac{g_{_B}^2}{2}\Big[\Big(g_{_V}^d\Big)^2+\Big(g_{_A}^d\Big)^2\Big]Br(Z^\prime\rightarrow l^+l^-)\;,
\label{zPrime-9}
\end{eqnarray}
and $w_{q}(s,m_{Z^\prime}^2)$ $\;$$(q=u,\;d)$are determined by the parton luminosities
$\frac{dL_{_{q\bar{q}}}}{ dm_{Z^\prime}^2}$, and thus only depend on the
collider energy $\sqrt{s}$ and $m_{Z^\prime}$.

By comparing the experimental limits of the LHC at $7\;{\rm GeV}$ to the theoretical
predictions of $c_{_u}-c_{_d}$ plane, the authors of Ref.~\cite{Accomando} obtained the
lower bound on $m_{Z^\prime}/g_{_B}\sim1730\;{\rm GeV}$ in the $U(1)_{B-L}$ extension
of the SM. This value is approximately equal to lower bound on $m_{Z^\prime}/g_{_B}$
of the sequential standard model (SSM). Although the authors of~\cite{Mzb} did not
specify a lower bound on $m_{Z^\prime}/g_{_B}$  in the $U(1)_{B-L}$ extension
of the SM with the updated LHC data. It is reasonable to assume that the lower bound on
$m_{Z^\prime}/g_{_B}\sim6\;{\rm TeV}$ in the $U(1)_{B-L}$ extension is approximately
equal to that in the SSM by the LHC data at $13$ TeV. Furthermore, CMS data
of the channel $pp\rightarrow Z^\prime\rightarrow e^+e^-$ at LHC of 13 TeV give the lower bound
on the $Z^\prime$ mass as $m_{Z^\prime}\ge4.72$ TeV, that of the channel
$pp\rightarrow Z^\prime\rightarrow\mu^+\mu^-$ give $m_{Z^\prime}\ge4.89$ TeV,
and the combination of two channel sets $m_{Z^\prime}\ge5.15$ TeV~\cite{Sirunyan}, respectively.
When $Z^\prime$ can decay into the superparticles in the supersymmetry $U(1)_{B-L}$
extension of the SM, the increasing of the total width of $Z^\prime$
decreases the lower bound on  the mass of $Z^\prime$.

So far the most stringent constraint on the $U(1)_{B-L}$ gauge boson parameters originates
from LEP2 results actually, which implies $M_{Z^\prime}/g_{_B}\geq 6\;{\rm{TeV}}$
at 95\% C.L. This bound is based on the assumption that $Z'$ dominantly decays to SM fermions
and is derived from the limit on the low energy four-fermion contact interactions induced by $Z^\prime$ exchanged diagram.
In the literature~\cite{Abdallah}, the authors also take this lower bound
for $Z^\prime$ mass to investigate the invisible decay of $Z^\prime$ in the B-LSSM.

\section{Matrix Elements and Cross Section \label{sec3}}

\begin{figure}
\setlength{\unitlength}{1mm}
\centering
\begin{minipage}[c]{0.9\textwidth}
\includegraphics[width=5in]{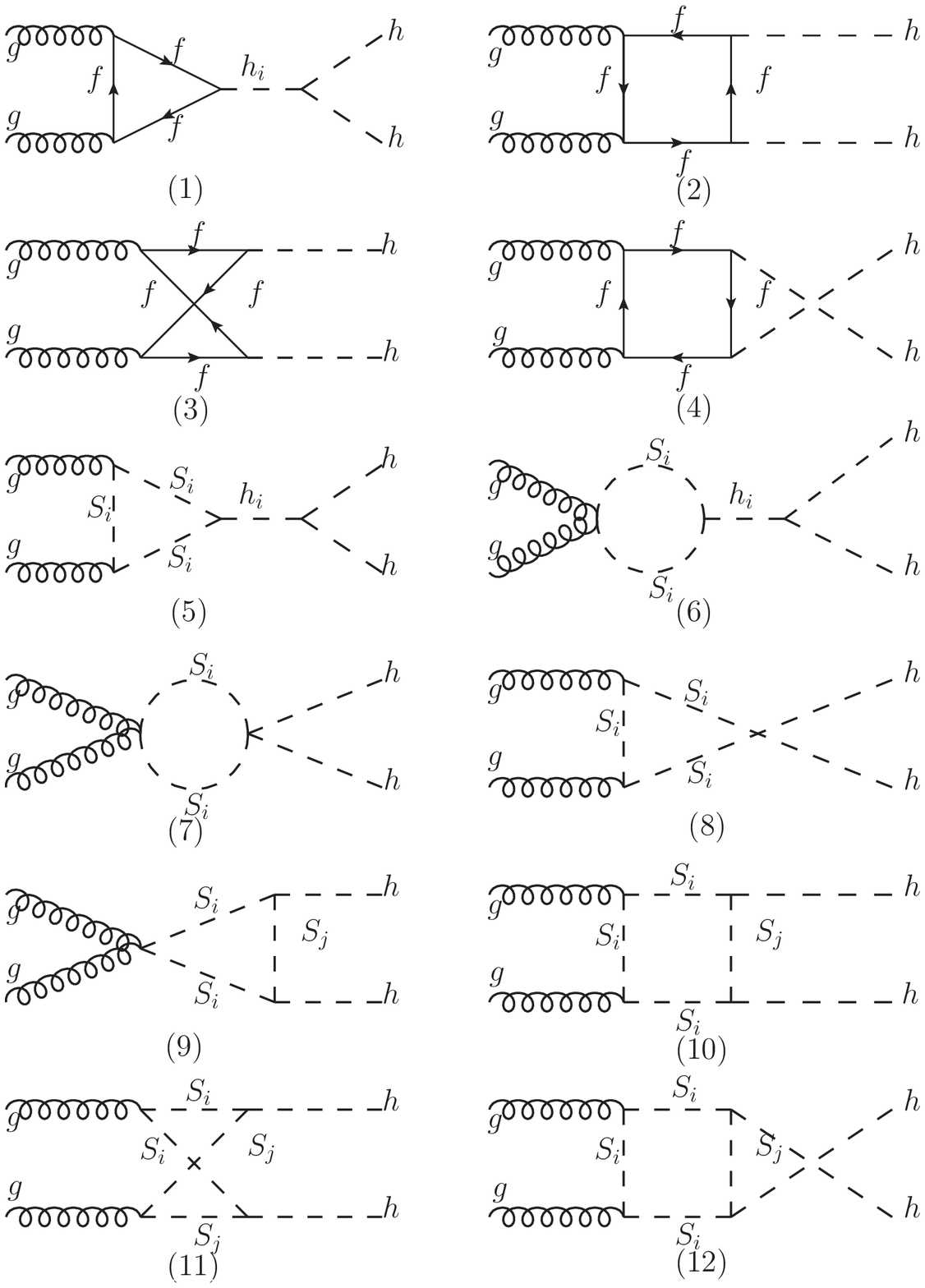}
\end{minipage}
\caption[]{Leading order Feynman diagrams for Higgs pair production in the B-LSSM at the LHC. $h_i$ are CP-even Higgs, $f$ are top and bottom quarks and $S_{i(j)}$ are stop and sbottom quarks.}
\label{Feynman diagrams}
\end{figure}

In this section, we analyze the pair production cross section of the lightest Higgs
$\sigma(pp\rightarrow gg \rightarrow hh)$ at $\sqrt{s}=14\:{\rm{TeV}}$  using the
analytical expressions for one-loop amplitudes of $g(p_1)g(p_2)\rightarrow h(p_3)h(p_4)$
in supersymmetric extensions of the SM. Feynman diagrams contributing to the $gg\rightarrow hh$ process
at LO are summarized in Fig.\ref{Feynman diagrams}. Some triangular diagrams contain information of trilinear
couplings among $CP$-even Higgs. As masses of scalar quarks all exceed ${\rm TeV}$ scale, the corrections
originating from the scalar quarks to the cross section of the Higgs pair production can be ignored safely.
Assuming that the Yukawa couplings of top and bottom quarks are determined through other experimental measurements,
one can analyze the correlations among the mass of the next-to-lightest
neutral Higgs $m_{_{h_2}}$ and the trilinear couplings $C_{_{hhh}},\;C_{_{h_2hh}}$
from the experimental data on the pair production cross section. Here $h_2$
denotes the next-to-lightest neutral Higgs, $C_{_{hhh}}$ denotes the trilinear coupling among
the lightest neutral Higgs, and $C_{_{h_2hh}}$ denotes the coupling among one next-to-lightest
neutral Higgs and two lightest neutral Higgs, respectively.
To ensure the correctness of our calculation,
we recover the SM results presented in Ref.~\cite{ quark-mssm} and the MSSM
results presented in Ref.~\cite{SM-higgs}.
In order to calculate the polarization cross section, we introduce the explicit
polarization vectors of the helicities ${(\lambda_1,\;\lambda_2)}$ for gluons as follows
\begin{eqnarray}
&&\epsilon^\mu_1(p_1,\lambda_1=\pm1)=\frac{1}{\sqrt{2}}(0,\mp1,-i,0)\;,\nonumber\\
&&\epsilon^\mu_2(p_2,\lambda_2=\pm1)=\frac{1}{\sqrt{2}}(0,\pm1,-i,0)\;.
\label{polarization vectors}
\end{eqnarray}
In the center-of-mass frame the momenta can be written in terms of the beam energy E and scattering angle $\theta$
\begin{eqnarray}
&&p^\mu_1=E(1,0,0,-1)\;,\nonumber\\
&&p^\mu_2=E(1,0,0,1)\;,\nonumber\\
&&p^\mu_3=E(1,-\sqrt{1-\frac{4m^2_h}{\hat{s}}} \sin\theta,0,-\sqrt{1-\frac{4m^2_h}{\hat{s}}} \cos\theta)\;,\nonumber\\
&&p^\mu_4=E(1,\sqrt{1-\frac{4m^2_h}{\hat{s}}} \sin\theta,0,\sqrt{1-\frac{4m^2_h}{\hat{s}}} \cos\theta)\;.
\label{momenta}
\end{eqnarray}
where $p_{1,2}$ are the incoming momenta of two gluons, and $p_{3,4}$
are the outgoing momenta of two lightest Higgs, respectively.
In addition, the Mandelstam variables are defined as
 \begin{eqnarray}
\hat{s}=(p_1+p_2)^2=(p_3+p_4)^2\;,\nonumber\\
\hat{t}=(p_1-p_3)^2=(p_2-p_4)^2\;,\nonumber\\
\hat{u}=(p_1-p_4)^2=(p_2-p_3)^2\;.
\label{Mandelstam variables}
\end{eqnarray}
The cross section of the subprocess $gg\rightarrow hh$ at LO can be written as
\begin{eqnarray}
&&{\hat{\sigma}}=\int^{\hat t_{max}}_{\hat t_{min}}\; d\hat{t} \;\frac{1}{4096\pi \hat{s}^2}
\;{(|\sum_{n}M^{(n)}_{++}|^2+|\sum_{n}M^{(n)}_{+-}|^2
+|\sum_{n}M^{(n)}_{--}|^2+|\sum_{n}M^{(n)}_{-+}|^2)}\;,\nonumber\\
\label{subprocess cross section}
\end{eqnarray}
where
\begin{eqnarray}
{\hat t_{min}}=(m^2_h-\frac{\hat{s}}{2})-\frac{1}{2}\;\sqrt{1-\frac{4m^2_h}{\hat{s}}}\;\hat{s}\;,\nonumber\\
{\hat t_{max}}=(m^2_h-\frac{\hat{s}}{2})+\frac{1}{2}\;\sqrt{1-\frac{4m^2_h}{\hat{s}}}\;\hat{s}\;.
\end{eqnarray}
Additionally $M_{{\lambda_1}{\lambda_2}}^{(n)}$ is the helicity amplitude
for pair production of the lightest neutral Higgs boson.
$CP$ conservation implies that $M_{++}=M_{--}$, $M_{+-}=M_{-+}$,
where the subscripts $\pm$ denote two independent helicities of the initial gluons.
The nonzero amplitudes from the Feynman diagrams in Fig.~\ref{Feynman diagrams}
are summarized in Eq.~(\ref{amplitudes}).

The differential cross section of $pp\rightarrow gg\rightarrow hh$ is
\begin{eqnarray}
\frac{d\sigma}{d\sqrt{\hat{s}}}=\frac{2\sqrt{\hat{s}}}{s}\:{\hat\sigma(gg-hh)}\:\frac{dL_{gg}}{d\tau}\;,
\label{differential cross section}
\end{eqnarray}
where $\tau=\hat{s}/s$, and
\begin{eqnarray}
\frac{dL_{gg}}{d\tau}=\int^1_\tau\;\frac{dx}{x}\;f_g(x,\mu_{_F})\;f_g(\frac{\tau}{x},\mu_{_F})\;.
\label{luminosity}
\end{eqnarray}
Here $f_g(x,\mu_{_F})$ is the parton distribution function of gluons,
$x$ is the relative momentum of the radiated gluon,
and $\mu_{_F}$ is the factorization scale, respectively.

The total cross section for the lightest neutral Higgs pair production
through $gg$ in $pp$ collision can be written as
\begin{eqnarray}
\sigma_{LO}(pp\rightarrow gg\rightarrow hh)=\int^1_{\tau_0}\:d\tau
\:{\hat\sigma(gg\rightarrow hh)}\:\frac{dL_{gg}}{d\tau}\;,
\label{cross section}
\end{eqnarray}
at the LO approximation. Here $\tau_0=(2m_h)^2/s$, and the loop integrals
are evaluated by the package LoopTools~\cite{Looptools}.

It is well known that the QCD corrections enhance the theoretical evaluation
of the cross section of the lightest Higgs pair production drastically.
The NLO QCD corrections~\cite{QCD-correction,QCD-correction1} to
the lightest Higgs pair production via gluon fusion have been
computed in the heavy-top limit (HTL)~\cite{QCD-correction}.
The NLO results for the gluon-fusion cross section are summarized
generally as~\cite{QCD-correction}
 \begin{eqnarray}
\sigma_{NLO}=\sigma_{LO}+\Delta\sigma_{virt}+\Delta\sigma_{gg}+\Delta\sigma_{gq}+\Delta\sigma_{q\bar{q}}\;,
\label{NLO cross section}
\end{eqnarray}
where the LO contribution to the cross section is given by Eq.(\ref{cross section}).
Other pieces are presented as
\begin{eqnarray}
&&\Delta\sigma_{virt}=\frac{(\alpha_s{(\mu_{_R})})}{\pi}\:\int^1_{\tau_0}\:d\tau\:
\frac{dL_{gg}}{d\tau}\:{\hat\sigma}_{LO}(\hat{s}=\tau{s})\:C_{virt}(\hat{s})\:,
\nonumber\\
&&\Delta\sigma_{ij}=\frac{(\alpha_s{(\mu_{_R})})}{\pi}\:\int^1_{\tau_0}\:d\tau\:
\frac{dL_{ij}}{d\tau}\:\int^1_{\frac{\tau_0}{\tau}}\:\frac{dz}{z}\:
{\hat\sigma}_{LO}(\hat{s}=z\tau{s})\:C_{ij}(\hat{s},z)\:.
\label{QCD correction}
\end{eqnarray}
where $\frac{dL_{ij}}{d\tau}$\:($i,j =g, q, \bar{q}$) denote the parton-parton
luminosities which are defined analogously to $\frac{dL_{gg}}{d\tau}$ in Eq.(\ref{luminosity}).
$q(x,{\mu_{_F}})$ is the quark parton distribution function , and
$z$ is the ratio between momentum of the radiated quark and that of the parent proton.
Furthermore the concrete expressions for $C_{ij}(\hat{s},z)$ ($ij=gg,\;gq,\;q\bar{q}$)
are presented in appendix~\ref{app1}.

\section{Numerical results\label{sec4}}

In the following we present some numerical results of the lightest Higgs pair production cross section for
the process $ pp\rightarrow gg \rightarrow hh $. The relevant SM inputs are $ m_{_t}=172.76\;{\rm{GeV}}$,
$m_{_b}=4.18\;{\rm{GeV}}$, $m_{_{\rm Z}}=90.19\;{\rm{GeV}}$, $m_{_{\rm W}}=80.385\;{\rm{GeV}}$ and
$\alpha_{em}(m_{_{\rm Z}})=1/128.9$, respectively. Choosing the renormalization scale $\mu_{_R}$ as the invariant mass of
the lightest Higgs pair, we adopt the CTEQ5 ~\cite{Cteq} to generate the parton distribution
function with the factorization scale $\mu_{_F}$. Furthermore the collision energy of the LHC is fixed to be $14\;{\rm{TeV}}$.
Taking $m_h=125\;{\rm{GeV}}$, we find that the theoretical prediction for
the Higgs pair production cross section in the SM is approximately $33.1\;{\rm{fb}}$,
which is consistent with result in ~\cite{B.Micco}.
To be consisted with the updated experimental data, we choose $M_{Z'}=5.2\;{\rm{TeV}}$
in our numerical analysis. Furthermore the result of Refs.~\cite{gb,gb1}
indicates an lower bound on the ratio between the $Z'$ mass and its gauge coupling
as $M_{Z'}/g_{_B}\geq 6\;{\rm{TeV}}$ at $99 $\%$ $ CL.
Under the above assumption on the mass of heavy neutral
vector boson, the scope of $g_{_B}$ is limited to $0<g_{_B}\leq0.8$.
In general, the Yukawa coupling $Y_b=\sqrt{2}m_{_b}/(\upsilon c_\beta)$
is smaller than one, so the parameter $\tan\beta$ should be approximatively smaller than 40,
and the large $\tan\beta$ has been excluded by the experimental data on
$Br(\bar{B}\rightarrow X_s\,\gamma)$ and $Br(B^0_s\rightarrow\mu^+\mu^-)$~\cite{tan}.
Additionally LHC experimental data~\cite{BLSSMB1} constrains the parameter $\tan\beta'<1.5$.
Considering the constraints of the experiments~\cite{PDG2},
we appropriately choose $M_{_{BB}}=M_{_{BB'}}=600\;\rm{GeV}$, $\mu^\prime=800\;\rm{GeV}$,
and $m_{\tilde{Q}}=m_{\tilde{U}}=m_{\tilde{D}}={\rm diag}(2,\;2,\;1.8)\;\rm{TeV}$, respectively.

Actually a natural configuration of the B-LSSM includes another relatively
light $CP$-even Higgs $h_2$, with mass $m_{h_2}\geq135\;{\rm{GeV}}$ besides the lightest
Higgs state with mass around $125\;{\rm GeV}$.
This fact was exploited in Refs.~\cite{next-to-lightest Higgs,next-to-lightest Higgs1,next-to-lightest Higgs2}
to explain potential Run I signal for another Higgs bosons, such as $h_2\rightarrow zz^\ast \rightarrow 4 \ell $~\cite{hzz},
$h_2\rightarrow rr$~\cite{hrr1,hrr2} and $h_2\rightarrow zr$~\cite{hzr} decay modes.
In the MSSM, two $CP$-even neutral Higgs are obtained through the mixing
between the real neutral components of two $SU(2)_L$ doublets.
It is well known that those radiative corrections from the third generation quarks
and scalar quarks drastically enhance the theoretical prediction on the mass of the
lightest Higgs. In the B-LSSM, four $CP$-even neutral Higgs are given through the mixing
among the real neutral components of two $SU(2)_L$ doublets and two singlets.
For this reason, there is not an upper limit on the mass of the
lightest Higgs at tree level.

\begin{figure}[!htbp]
\setlength{\unitlength}{1mm}
\centering
\begin{minipage}[c]{0.5\textwidth}
\includegraphics[width=3.2in]{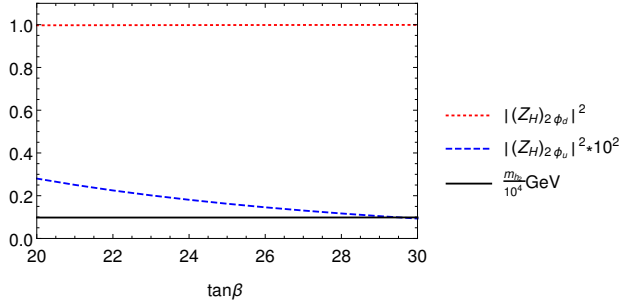}
\end{minipage}
\caption[]{ The mass of
the next-to-lightest $CP$-even Higgs and square of absolute value of the mixing elements
versus $\tan\beta$ in the MSSM.}
\label{num1}
\end{figure}

\begin{figure}[!htbp]
\setlength{\unitlength}{1mm}
\centering
\begin{minipage}[c]{0.5\textwidth}
\includegraphics[width=3.2in]{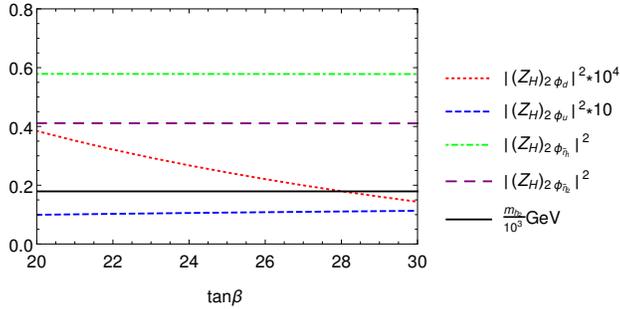}
\end{minipage}
\caption[]{ The mass of
the next-to-lightest $CP$-even Higgs and square of absolute value of the mixing elements
versus $\tan\beta$ in the B-LSSM.}
\label{num2}
\end{figure}

In the numerical analysis we choose those parameters so that the corresponding
theoretical prediction of mass of the lightest Higgs coincides with the experimental data
within 3 standard deviations. In the MSSM, the ratio $\tan\beta$ between two doublet nonzero VEVs
and the soft breaking parameters $B_{_\mu}$, $A_{_t}$, and $A_{_b}$ strongly affect the theoretical
predictions on the masses of $CP$-even Higgs together with their mixing.
In the B-LSSM, the ratio $\tan\beta^\prime$ between two singlet nonzero VEVs and additional
gauge couplings $g_{_{YB}}$ and $g_{_{B}}$ also affect the theoretical
predictions on the masses of $CP$-even Higgs besides those aforementioned in the MSSM.
The present experimental data of the lightest Higgs mass
sets a strong constraint on the parameter space of various extensions
of the SM. In order to fit the experimental data of the lightest Higgs,
we assume $B_{_\mu}=0.5\;{\rm TeV}^2$, $\mu=800\;{\rm GeV}$, and $A_{_t}=A_{_b}=0.6\;{\rm TeV}$,
respectively in the MSSM. The lightest Higgs around $125\;{\rm GeV}$ is mainly composed by component
originating from the real part of neutral $\phi_{_u}$, and little component originating
from the real part of neutral $\phi_{_d}$. Under our assumption on the parameter space
of the MSSM, we plot the mass of the next-to-lightest (heaviest)$CP$-even Higgs and
square of absolute value of the mixing elements versus $\tan\beta$ in Fig.\ref{num1}.
The numerical results indicate that the next-to-lightest (heaviest) $CP$-even Higgs
is given the mass around $1\;{\rm TeV}$, and composed mainly by the component originating from the
real part of neutral $\phi_{_d}$. In the B-LSSM, the lightest Higgs around $125\;{\rm GeV}$
is also composed dominantly by component originating from the real part of neutral $\phi_{_u}$,
and little component originating from the real part of neutral $\phi_{_d}$, together with
that of two singlets $\phi_{\tilde{{\eta}}_1},\;\phi_{\tilde{{\eta}}_2}$. To fit the lightest Higgs
with mass around $125\;{\rm GeV}$, we take $B_{_\mu}=B_{_{\mu^\prime}}=0.5\;{\rm TeV}^2$,
$\mu=600\;{\rm GeV}$, $A_{_t}=A_{_b}=1.6\;{\rm TeV}$, $-g_{_{YB}}=g_{_B}=0.4$, $\mu^\prime=800\;{\rm GeV}$, and
$\tan\beta^\prime=1.2$, respectively.
Adopting the assumption above on the parameter space of the B-LSSM, we plot the mass of
the next-to-lightest $CP$-even Higgs and square of absolute value of the mixing elements
versus $\tan\beta$ in Fig.\ref{num2}. The numerical results indicate that
the next-to-lightest $CP$-even Higgs obtains the mass around $180\;{\rm GeV}$, and
is composed mainly by the components originating from the real part of two singlets
$\phi_{\tilde{{\eta}}_1},\;\phi_{\tilde{{\eta}}_2}$.

\begin{figure}[!htbp]
\setlength{\unitlength}{1mm}
\centering
\begin{minipage}[c]{0.5\textwidth}
\includegraphics[width=3in]{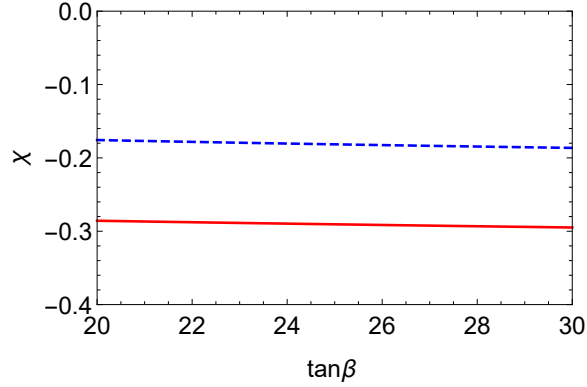}
\end{minipage}
\caption[]{ In the MSSM, the red solid and blue dashed lines represent $\chi=\frac{C{_{hhh}}-C^{(0)}{_{hhh}}}{C^{(0)}{_{hhh}}}$
and $\chi=\frac{C{_{{h_2}hh}}-C^{(0)}{_{{h_2}hh}}}{C^{(0)}{_{{h_2}hh}}}$versus $\tan\beta$, respectively.
$C^{(0)}{_{{h_i}hh}}$ denotes the $CP$-even Higgs trilinear couplings at the tree level.
 $C{_{{h_i}hh}}$ denotes the $CP$-even Higgs trilinear couplings including radiative corrections.}
\label{num3}
\end{figure}

\begin{figure}[!htbp]
\setlength{\unitlength}{1mm}
\centering
\begin{minipage}[c]{0.5\textwidth}
\includegraphics[width=3in]{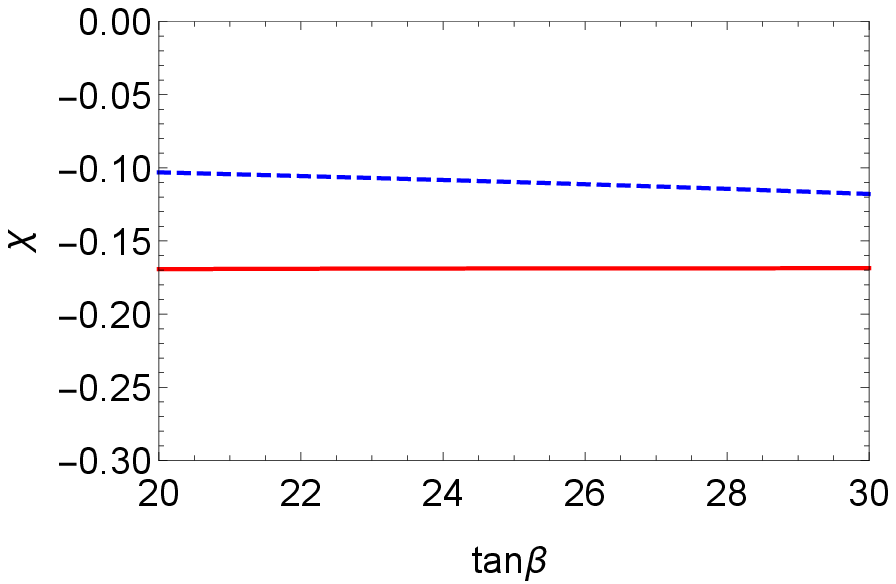}
\end{minipage}
\caption[]{ In the B-LSSM, the red solid and blue dashed lines represent $\chi=\frac{C{_{hhh}}-C^{(0)}{_{hhh}}}{C^{(0)}{_{hhh}}}$
and $\chi=\frac{C{_{{h_2}hh}}-C^{(0)}{_{{h_2}hh}}}{C^{(0)}{_{{h_2}hh}}}$ versus $\tan\beta$, respectively.
$C^{(0)}{_{{h_i}hh}}$ denotes the $CP$-even Higgs trilinear couplings at the tree level.
 $C{_{{h_i}hh}}$ denotes the $CP$-even Higgs trilinear couplings including radiative corrections.}
\label{num4}
\end{figure}

Besides enhancement of the theoretical prediction for the mass of the lightest $CP$-even Higgs,
the radiative corrections also modify the trilinear couplings among Higgs drastically.
Taking $B_{_\mu}=0.5\;{\rm TeV}^2$, $\mu=800\;{\rm GeV}$, and $A_{_t}=A_{_b}=0.6\;{\rm TeV}$
in the MSSM, we present the relative radiative corrections to trilinear coupling of
the lightest Higgs and that between the lightest Higgs and the next-to-lightest Higgs versus
$\tan\beta$ in Fig.\ref{num3}, respectively. The radiative contributions reduce trilinear coupling of
the lightest Higgs about $16\%$, and reduce that between the lightest Higgs and
the next-to-lightest Higgs $10\%$, separately. Similarly assuming $B_{_\mu}=B_{_{\mu^\prime}}=0.5\;{\rm TeV}^2$,
$\mu=600\;{\rm GeV}$, $\mu^\prime=800\;{\rm GeV}$, $A_{_t}=A_{_b}=1.6\;{\rm TeV}$, $-g_{_{YB}}=g_{_B}=0.4$, and
$\tan\beta^\prime=1.2$ in the B-LSSM, we draw the relative radiative corrections to trilinear coupling of
the lightest Higgs and that between the lightest Higgs and the next-to-lightest Higgs versus
$\tan\beta$ in Fig.\ref{num4} respectively. Radiative contributions modify trilinear coupling of
the lightest Higgs about $30\%$, and decrease that among the lightest Higgs and
the next-to-lightest Higgs $18\%$, separately.

\begin{figure}[!htbp]
\setlength{\unitlength}{1mm}
\centering
\begin{minipage}[c]{0.5\textwidth}
\includegraphics[width=3in]{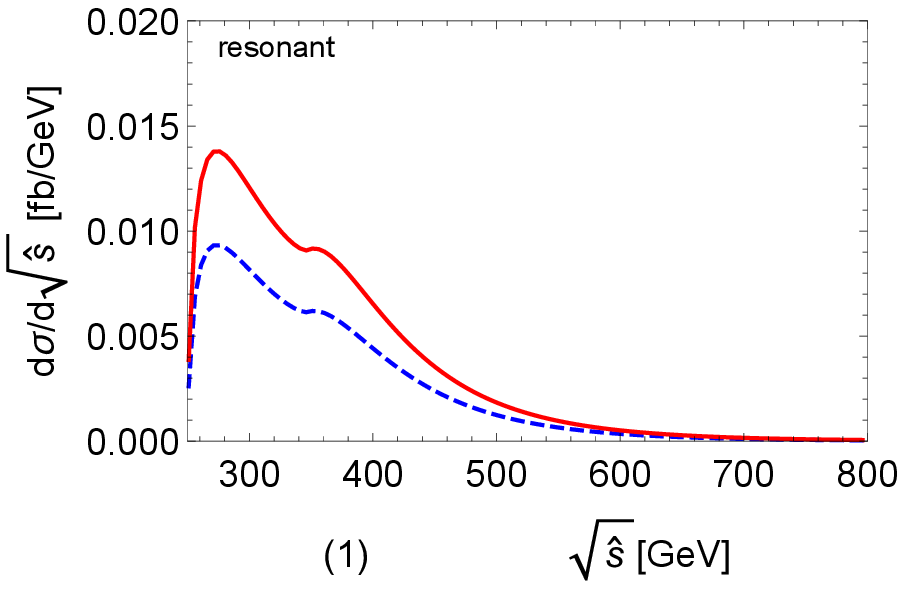}
\end{minipage}%
\begin{minipage}[c]{0.5 \textwidth}
\includegraphics[width=3in]{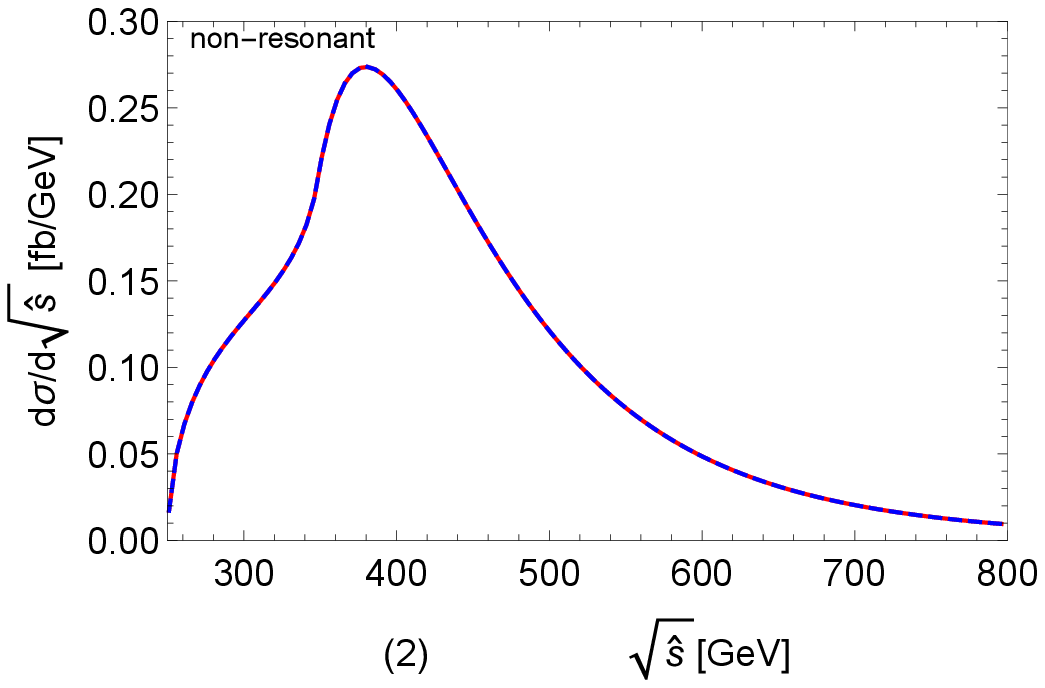}
\end{minipage}
\begin{minipage}[c]{0.5\textwidth}
\includegraphics[width=3in]{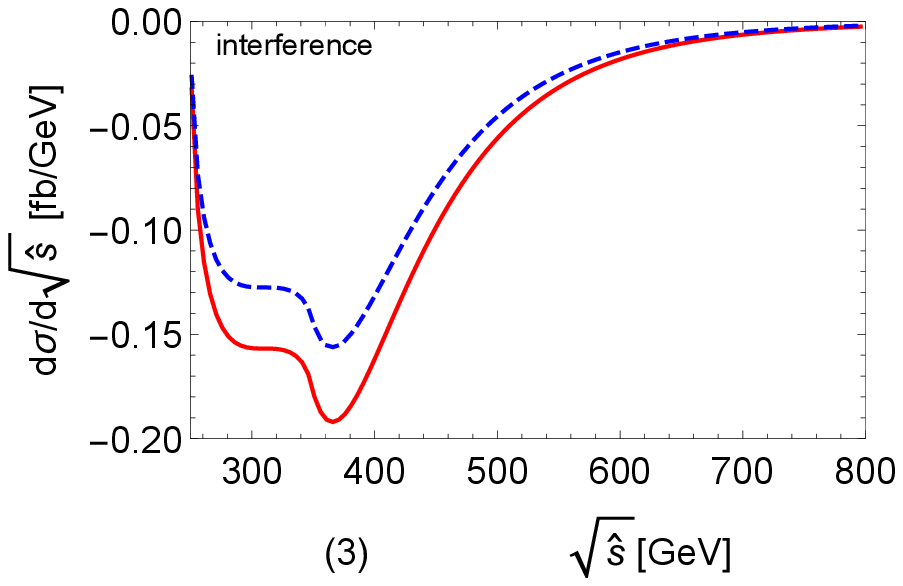}
\end{minipage}%
\begin{minipage}[c]{0.5\textwidth}
\includegraphics[width=3in]{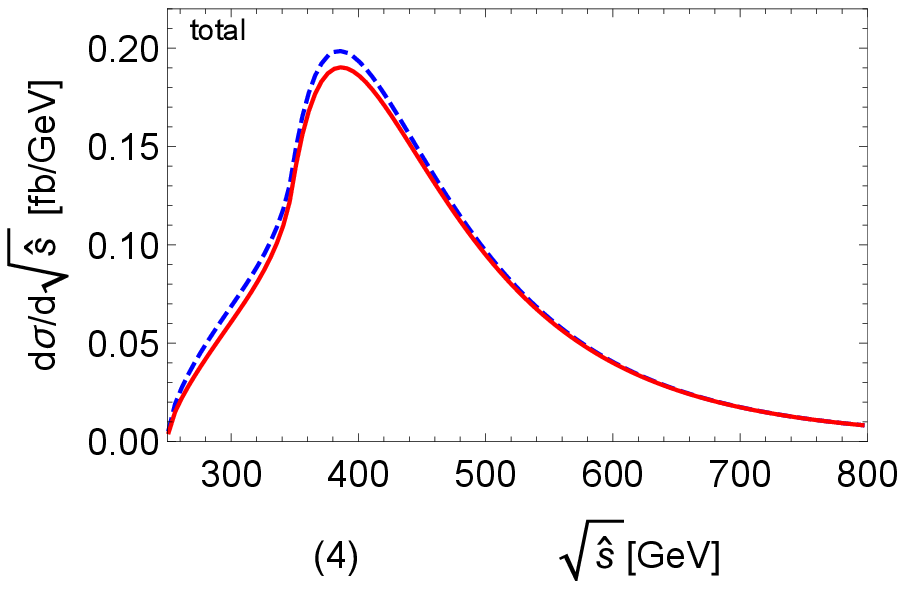}
\end{minipage}
\caption[]{ The resonant, non-resonant, interference, and the total differential cross section versus
$\sqrt{\hat{s}}$ in the MSSM. Where the blue dashed line represents the theoretical prediction
on the differential cross section including the radiative corrections, while the red solid line does not.}
\label{num5,6,7,8}
\end{figure}

Including the NLO QCD corrections, we present the cross section of the lightest Higgs pair
production as $\sqrt{s}=14\;{\rm{TeV}}$ at the LHC.
In order to present our results transparently, we separately plot the differential
cross section according to resonant, non-resonant, interference and total contribution.
At the LO, the differential cross section is
\begin{eqnarray}
\frac{d\sigma}{d\sqrt{\hat{s}}}=\frac{2\sqrt{\hat{s}}}{s}\Big\{{\hat\sigma_{res}}+{\hat\sigma_{nr}}
+{\hat\sigma_{int}}\Big\}\frac{dL_{gg}}{d\tau}\;,
\end{eqnarray}
with
\begin{eqnarray}
&&{\hat\sigma_{res}}=\frac{1}{4096\pi^2}\int^{\hat t_{max}}_{\hat t_{min}}\; d\hat{t}\;\frac{|M_{res}|^2}{\hat{s}^2}
\;,\nonumber\\
&&{\hat\sigma_{nr}}=\frac{1}{4096\pi^2}\int^{\hat t_{max}}_{\hat t_{min}}\; d\hat{t}\;\frac{|M_{nr}|^2}{\hat{s}^2}
\;,\nonumber\\
&&{\hat\sigma_{int}}=\frac{1}{4096\pi^2}\int^{\hat t_{max}}_{\hat t_{min}}\; d\hat{t}\;\frac{M_{int}}{\hat{s}^2}\;.
\end{eqnarray}
Here
\begin{eqnarray}
&&\mid M_{res}\mid^2=2(M^{(1)}_{++}+M^{(5+6)}_{++})^2
\;,\nonumber\\
&&\mid M_{nr}\mid^2=2(\mid\sum_{n}M^{(n)}_{++}- (M^{(1)}_{++}+M^{(5+6)}_{++})\mid^2+\mid\sum_{n}M^{(n)}_{+-}\mid^2)
\;,\nonumber\\
&&M_{int}=2Re(2(M^{(1)}_{++}+M^{(5+6)}_{++})*{2(\sum_{n}M^{(n)}_{++}- (M^{(1)}_{++}+M^{(5+6)}_{++}))}^\ast)\;.
\label{resonant, non-resonant, and interference amplitudes }
\end{eqnarray}
Actually, the resonant amplitude $M_{res}$ originates from the diagrams $(1),\;(5),\;(6)$ in Fig.\ref{Feynman diagrams}
where the contributions of last two diagrams are suppressed by the ${\rm TeV}$ masses of squarks.
In other words, the resonant amplitude $M_{res}$ strongly depends on the trilinear couplings among $CP$-even Higgs.
The non-resonant amplitude $M_{nr}$ includes all other Feynman diagrams, and the interference amplitude
$M_{int}$ corresponds to the interference between resonant and non-resonant amplitude
which also depends on the aforementioned trilinear couplings of Higgs.
At the NLO approximation, the QCD corrections should be allotted among three pieces appropriately.
Choosing $\tan\beta=24$, $B_{_\mu}=0.5\;{\rm TeV}^2$, $\mu=800\;{\rm GeV}$, and $A_{_t}=A_{_b}=0.6\;{\rm TeV}$
in the MSSM, we present the resonant, non-resonant, interference and total contributions to the
differential cross sections versus the available centre-of-mass energy $\sqrt{\hat{s}}$ of the parton-parton
in Fig.\ref{num5,6,7,8}(1), Fig.\ref{num5,6,7,8}(2),
Fig.\ref{num5,6,7,8}(3), and Fig.\ref{num5,6,7,8}(4),
respectively. The numerical results indicate that the radiative corrections to the trilinear
couplings of $CP$-even Higgs reduce
the absolute values of the resonant and interference contributions of the differential cross sections
about $20\%$ as the centre-of-mass energy of the parton-parton $\sqrt{\hat{s}}\le450\;{\rm GeV}$,
and do not affect the non-resonant piece of the differential cross sections because the non-resoant
contribution does not depend on the trilinear couplings of the $CP$-even Higgs.

\begin{figure}[!htbp]
\setlength{\unitlength}{1mm}
\centering
\begin{minipage}[c]{0.5\textwidth}
\includegraphics[width=3in]{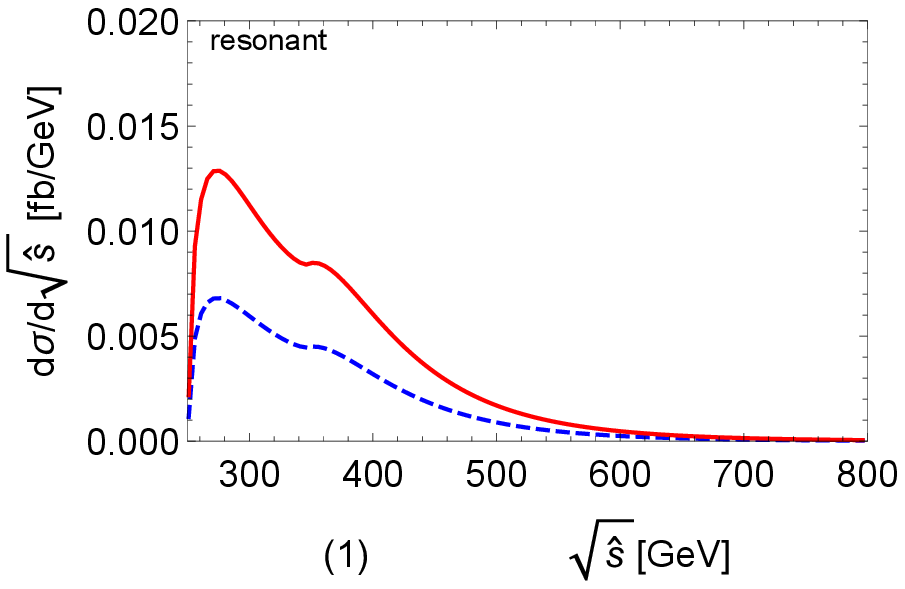}
\end{minipage}%
\begin{minipage}[c]{0.5\textwidth}
\includegraphics[width=3in]{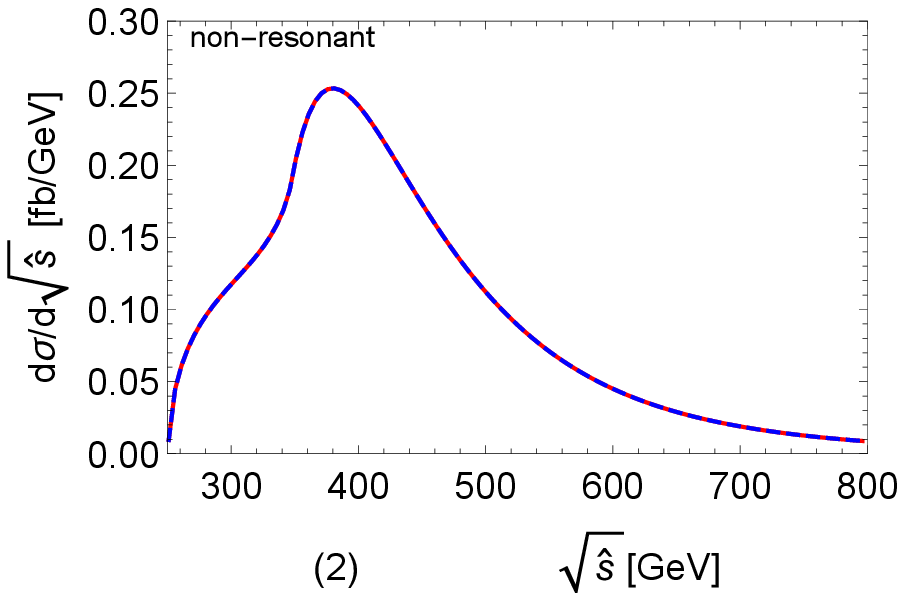}
\end{minipage}
\begin{minipage}[c]{0.5\textwidth}
\includegraphics[width=3in]{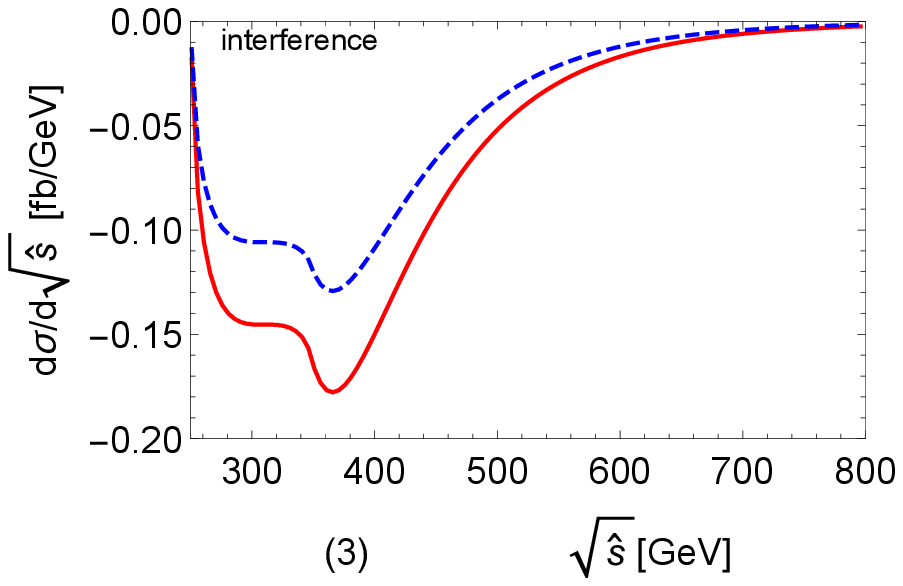}
\end{minipage}%
\begin{minipage}[c]{0.5\textwidth}
\includegraphics[width=3in]{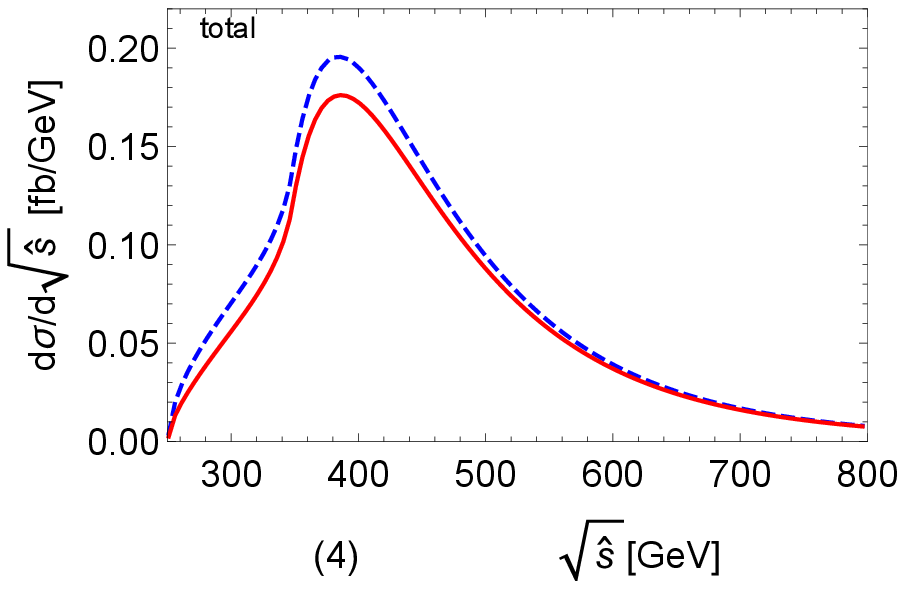}
\end{minipage}
\caption[]{ The resonant, non-resonant, interference, and the total differential
cross section versus $\sqrt{\hat{s}}$  in the B-LSSM.
Where the blue dashed line represents the theoretical prediction
on the differential cross section including the radiative corrections,
while the red solid line does not.}
\label{num9,10,11,12}
\end{figure}

Taking $\tan\beta=25$, $\tan\beta^\prime=1.2$, $g_{_B}=-g_{_{YB}}=0.4$,
$B_{_\mu}=B_{_{\mu^\prime}}=0.5\;{\rm TeV}^2$, $\mu=800\;{\rm GeV}$,
and $A_{_t}=A_{_b}=1.6\;{\rm TeV}$ in the B-LSSM, we present the resonant, non-resonant,
 interference and total contributions to the
differential cross sections versus the available centre-of-mass energy $\sqrt{\hat{s}}$ of the parton-parton
in Fig.\ref{num9,10,11,12}(1), Fig.\ref{num9,10,11,12}(2),
 Fig.\ref{num9,10,11,12}(3), and Fig.\ref{num9,10,11,12}(4),
respectively. The numerical results indicate that the radiative corrections
to the trilinear couplings of $CP$-even Higgs reduce
the absolute values of the resonant and interference pieces of the differential cross sections
about $40\%$ as the centre-of-mass energy of the parton-parton $\sqrt{\hat{s}}\le450\;{\rm GeV}$,
and do not affect the non-resonant piece of the differential cross sections.
At the NLO approximation, the resonant contribution is smaller than the non-resonant contribution,
thus the dominant contributions originate from the non-resonant sector and the destructive
interference between the resonant and non-resonant sectors.
Including the radiative corrections to the trilinear couplings of $CP$-even Higgs, the non-resonant
contribution remains invariantly, the resonant contribution decreases slightly and the interference
contribution decreases obviously. Therefore, the correction to the cross section originates mainly from the
interference sector which depends on the trilinear couplings of Higgs sensitively.
In addition, the radiative corrections to the $CP$-even Higgs trilinear
couplings enhance the cross section.

With the above assumptions on the parameter space of the MSSM and B-LSSM,
the relative correction to the differential cross section from the radiative corrections
to the trilinear couplings of $CP$-even Higgs in the MSSM is smaller than that in the B-LSSM
because the radiative correction to the trilinear couplings of $CP$-even
Higgs in the MSSM is significantly smaller than that in the B-LSSM.

\begin{figure}[!htbp]
\setlength{\unitlength}{1mm}
\centering
\begin{minipage}[c]{0.5\textwidth}
\includegraphics[width=3in]{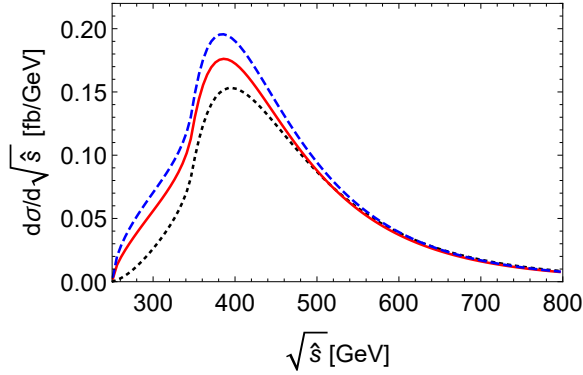}
\end{minipage}
\caption[]{ The differential cross section $\frac{d\sigma}{d\sqrt{\hat{s}}}$ versus $\sqrt{\hat{s}}$ in
the SM (dotted), MSSM (blue dashed) and B-LSSM (red solid).}
\label{num11}
\end{figure}

Furthermore, we present the differential cross section $\frac{d\sigma}{d\sqrt{\hat{s}}}$
versus $\sqrt{\hat{s}}$ in Fig.\ref{num11}, where the dotted line represents
the theoretical evaluation of the SM, the blue dashed line represents that of the MSSM, and the
red solid line represents that of the B-LSSM, respectively.
Along with the increasing of $\sqrt{\hat{s}}$ the differential cross section increases
as $250\;{\rm{GeV}}<\sqrt{\hat{s}}<400\;{\rm{GeV}}$, and decreases steeply
as $\sqrt{\hat{s}}>400\;{\rm{GeV}}$. Actually, the peak around $\sqrt{\hat{s}}\sim400\;{\rm{GeV}}$ is
caused by the interference effect of the resonant and non-resonant sectors. Choosing $m_h=125\;{\rm{GeV}}$,
we find that the theoretical prediction on the lightest Higgs pair
production cross section is about $33.1\;{\rm{fb}}$ in the SM, $37.9\;{\rm{fb}}$ in the B-LSSM and
$40.5\;{\rm{fb}}$ in the MSSM, respectively. The theoretical predictions of Higgs pair production
in the B-LSSM and that in the MSSM differ significantly from that in the SM.

\begin{figure}[!htbp]
\setlength{\unitlength}{1mm}
\centering
\begin{minipage}[c]{0.5\textwidth}
\includegraphics[width=3in]{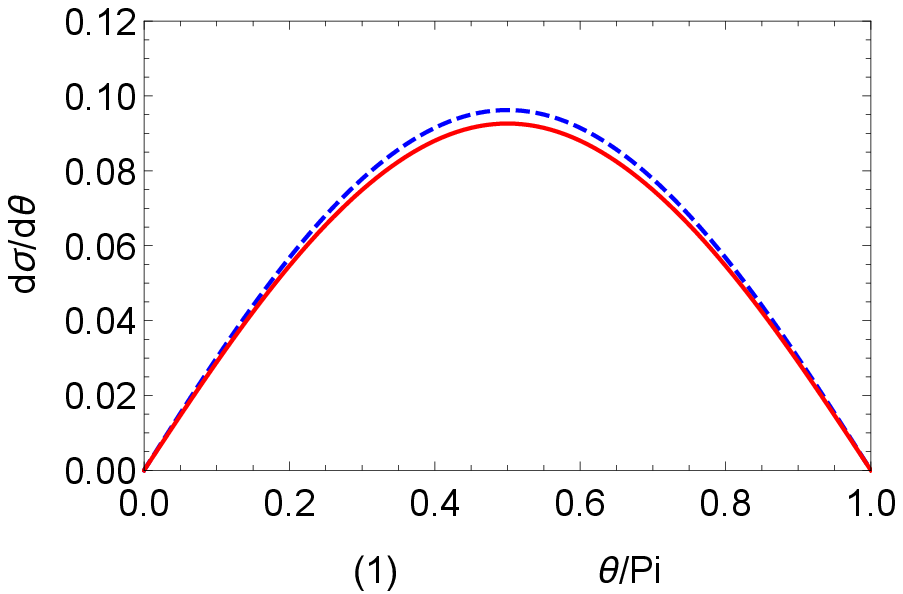}
\end{minipage}%
\begin{minipage}[c]{0.5\textwidth}
\includegraphics[width=3in]{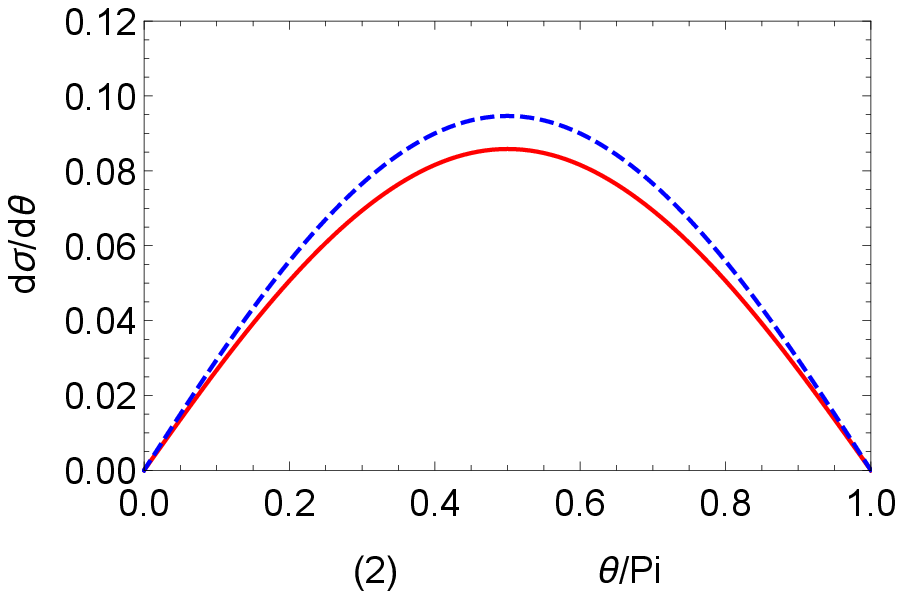}
\end{minipage}
\caption[]{The differential cross section $\frac{d\sigma}{d\theta}$ versus $\frac{\theta}{\pi}$ in the MSSM and B-LSSM.
The red solid line represents the theoretical prediction
on the differential cross section at NLO. The blue dashed line represents
the theoretical prediction on the differential cross section including the radiative corrections.}
\label{num14,15}
\end{figure}

Choosing $\sqrt{\hat{s}}=400\;{\rm{GeV}}$, $\tan\beta=24$, $B_{_\mu}=0.5\;{\rm TeV}^2$,
$\mu=800\;{\rm GeV}$, and $A_{_t}=A_{_b}=0.6\;{\rm TeV}$ in the MSSM,
we present the angle distribution of the differential cross section in
Fig.\ref{num14,15} (1).
Similarly taking $\tan\beta=25$, $\tan\beta^\prime=1.2$, $g_{_B}=-g_{_{YB}}=0.4$,
$B_{_\mu}=B_{_{\mu^\prime}}=0.5\;{\rm TeV}^2$, $\mu=800\;{\rm GeV}$,
and $A_{_t}=A_{_b}=1.6\;{\rm TeV}$ in the B-LSSM, we present the angle
distribution of the differential cross section in Fig.\ref{num14,15} (2).
Obviously, when $\theta=\frac{\pi}{2}$, the differential cross section reaches it maximum,
and when $\theta=0$ and $\pi$, the differential cross section reaches it minimum.

\begin{figure}[!htbp]
\setlength{\unitlength}{1mm}
\centering
\begin{minipage}[c]{0.5\textwidth}
\includegraphics[width=3in]{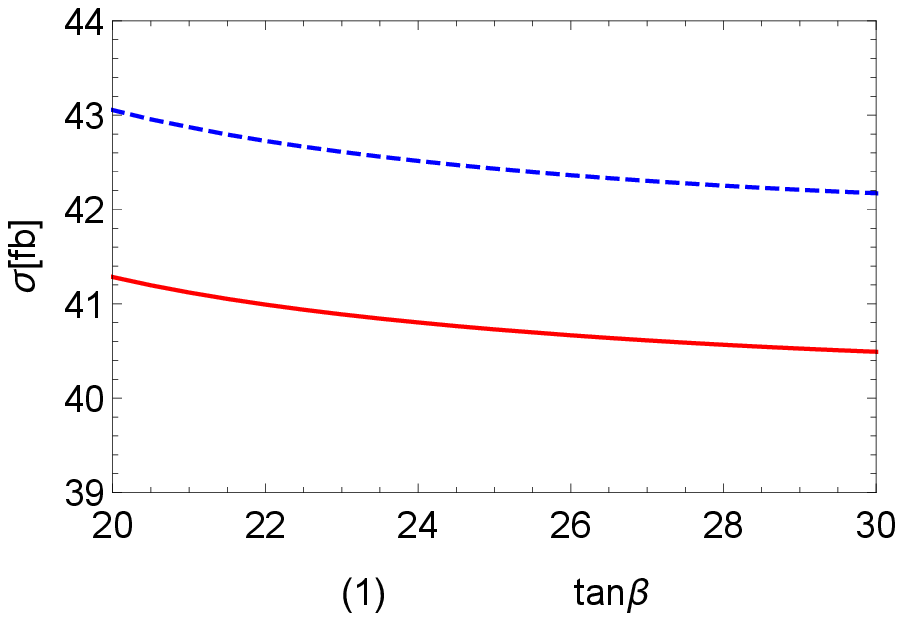}
\end{minipage}%
\begin{minipage}[c]{0.5\textwidth}
\includegraphics[width=3in]{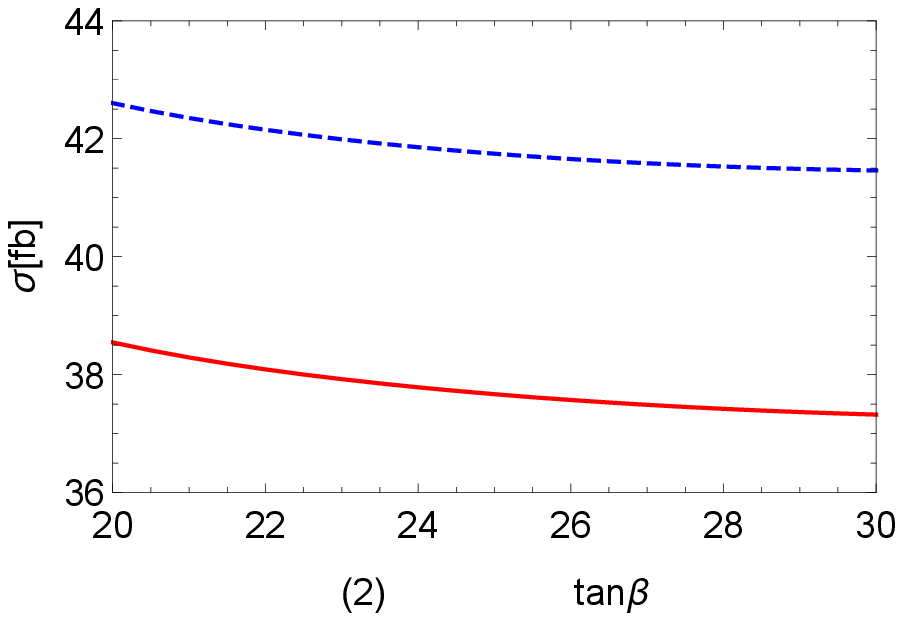}
\end{minipage}
\caption[]{ The total cross section $\sigma$ versus $\tan\beta$ in the MSSM
and B-LSSM, respectively. Where the red solid line represents the theoretical prediction
on the total cross section at NLO, the blue dashed line represents the theoretical prediction
on the total cross section including the radiative corrections.}
\label{num16,17}
\end{figure}

In order to satisfy the experimental constraints
mentioned above, we choose $B_{_\mu}=0.5\;{\rm TeV}^2$, $\mu=800\;{\rm{GeV}}$ and $A_t=A_b=0.6\;{\rm TeV}$
in the MSSM, and choose $\tan\beta'=1.2$, $g_{_B}=0.4$, $g_{_{YB}}=-0.4$, $B_{_\mu}=B_{_{\mu^\prime}}=0.5\;{\rm TeV}^2$,
$\mu=600\;{\rm{GeV}}$ and $A_t=A_b=1.6\;{\rm TeV}$ in the B-LSSM, respectively.
With those assumptions on the parameter space,
we plot the total cross section of the MSSM versus $\tan\beta$ in Fig.\ref{num16,17} (1)
and that of the B-LSSM versus $\tan\beta$ in Fig.\ref{num16,17} (2), respectively.
With the increasing of $\tan\beta$, the total cross section decreases mildly.
In the MSSM, the radiative correction to the theoretical prediction of the total cross section
is about $4\%$, while in the B-LSSM the radiative correction exceeds $11\%$ roughly.

\begin{figure}[!htbp]
\setlength{\unitlength}{1mm}
\centering
\begin{minipage}[c]{0.5\textwidth}
\includegraphics[width=3in]{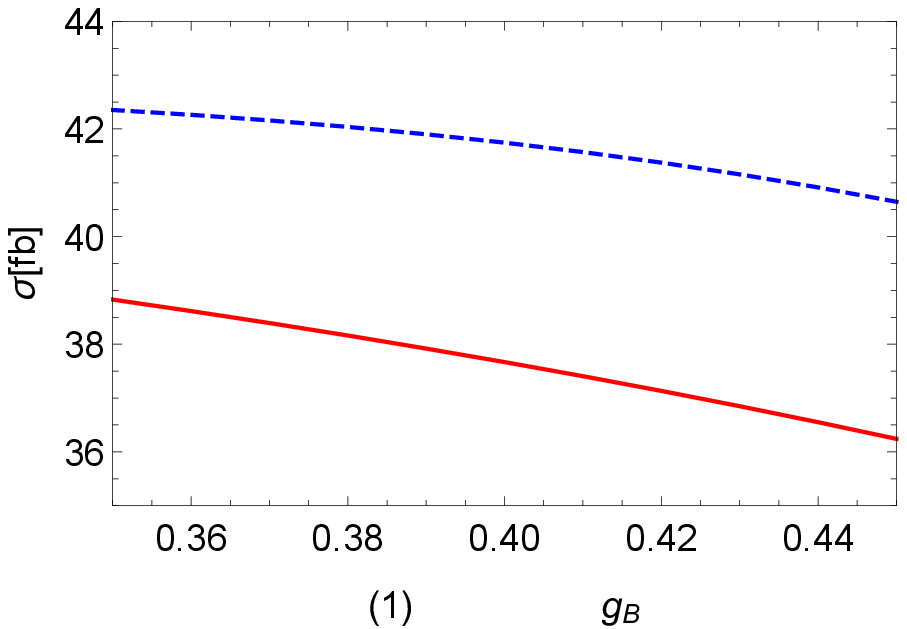}
\end{minipage}
\begin{minipage}[c]{0.5\textwidth}
\includegraphics[width=3in]{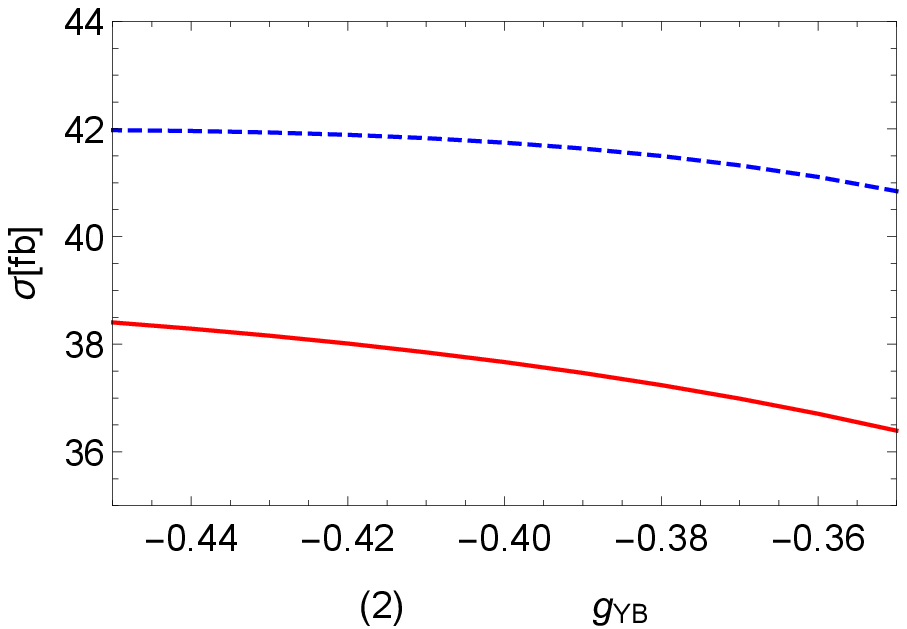}
\end{minipage}
\begin{minipage}[c]{0.5\textwidth}
\includegraphics[width=3in]{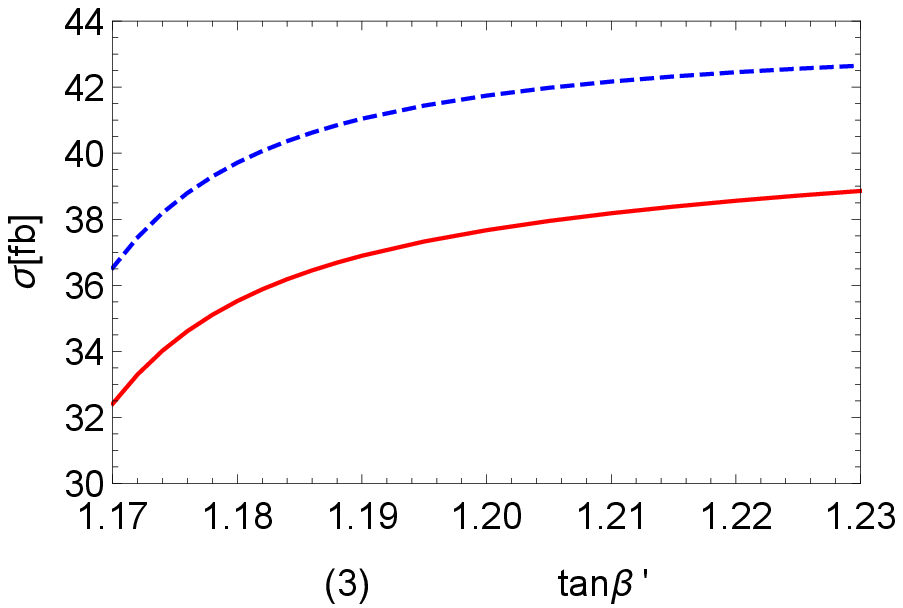}
\end{minipage}
\caption[]{ The total cross section $\sigma$ versus $g_{_B}$, $g_{_{YB}}$ and $ \tan\beta'$, respectively.
Where the red solid line represents the theoretical prediction
on the total cross section at NLO, the blue dashed line represents the theoretical prediction
on the total cross section including the radiative corrections.}
\label{num18,19,20}
\end{figure}

Finally we study the influence on the total cross section of the parameters
$g_{_B}$, $g_{_{YB}}$ and $\tan\beta'$ in the B-LSSM. Taking
$\tan\beta=25$, $\tan\beta'=1.2$, $g_{_{YB}}=-0.4$,
$B_{_\mu}=B_{_{\mu^\prime}}=0.5\;{\rm TeV}^2$, $\mu=800\;{\rm GeV}$,
and $A_{_t}=A_{_b}=1.6\;{\rm TeV}$ in the B-LSSM, we draw total cross section
versus the gauge coupling $g_{_B}$ in Fig.\ref{num18,19,20} (1).
With increasing of the coupling $g_{_B}$, the evaluation of total cross
section decreases quickly.
Similarly selecting $\tan\beta=25$, $\tan\beta'=1.2$, $g_{_{B}}=0.4$,
$B_{_\mu}=B_{_{\mu^\prime}}=0.5\;{\rm TeV}^2$, $\mu=800\;{\rm GeV}$,
and $A_{_t}=A_{_b}=1.6\;{\rm TeV}$, we draw total cross section versus the gauge
coupling $g_{_{YB}}$ in Fig.\ref{num18,19,20} (2). With the increasing of the coupling
$g_{_{YB}}$, the total cross section decreases mildly.
In Fig.\ref{num18,19,20} (3), we plot total cross section changing with
$\tan\beta'$ as $\tan\beta=25$, $g_{_{YB}}=-0.4$, $g_{_{B}}=0.4$,
$B_{_\mu}=B_{_{\mu^\prime}}=0.5\;{\rm TeV}^2$, $\mu=800\;{\rm GeV}$,
and $A_{_t}=A_{_b}=1.6\;{\rm TeV}$. With increasing of
$\tan\beta^\prime$, the evaluation of total cross section increases steeply.

\section{Summary\label{sec5}}

In this paper, we discuss the pair production of the lightest neutral Higgs in supersymmetric
extensions of the SM. The lightest neutral Higgs pair is produced dominantly in
$pp\rightarrow gg\rightarrow hh$ through the loop-induced gluon fusion mechanism.
We analyze theoretical evaluations of the lightest neutral Higgs pair production cross section,
and investigate the radiative corrections to the
trilinear couplings of $CP$-even Higgs. We find that the relatively
radiative corrections to the trilinear coupling of the lightest neutral Higgs is about $-30\%$ in
the B-LSSM and $-16\%$ in the MSSM. At the NLO approximation, the dominant
contributions to the cross section originate from the non-resonant sector and
the destructive interference between the resonant and non-resonant sectors.
The radiative correction to the trilinear couplings
of $CP$-even Higgs modifies the theoretical predictions on the cross section.
In the B-LSSM, the relative correction to the cross section is about 11\%, while in the MSSM, the relative
correction to the cross section is only about 4\%. Furthermore, we also discuss the effect
of some new parameters on the cross section in the B-LSSM, such as $g_{_B}, g_{_{YB}}$ and $\tan\beta'$.
We find that the theoretical prediction for the production
cross section of $pp\rightarrow gg\rightarrow hh$ depends on these parameters sensitively.

\begin{acknowledgments}
\indent\indent
We are very grateful to teachers Shu-Min Zhao and Yu-Shu Song from Hebei University, for giving us some useful discussions.
The work has been supported by the National Natural Science Foundation of China (NNSFC) with Grants No. 12075074
No. 11535002,  the youth top-notch talent support program of the Hebei Province,
and Midwest Universities Comprehensive Strength Promotion project.
\end{acknowledgments}

\appendix
\section{$2\times2$ submatrices\label{app-A}}
\indent\indent
The $2\times2$ submatrices in Eq.(\ref{mh-tree-level}) are formulated respectively as
\begin{eqnarray}
&&\Big[m_{_h}^2\Big]_{_{\phi\phi}}=\left(\begin{array}{cc}
\frac{1}{4}g^2\upsilon^2 c_{\beta}^2 +\rm{Re}(B_\mu) t_\beta,\;\; & -\frac{1}{4}g^2\upsilon^2 s_\beta c_{\beta}-\rm{Re}(B_\mu)
\\-\frac{1}{4}g^2\upsilon^2 s_\beta c_{\beta}-\rm{Re}(B_\mu),\;\; & \frac{1}{4}g^2\upsilon^2 s_{\beta}^2+\rm{Re}(B_\mu)/t_\beta\end{array}\right)
\;,\nonumber\\
&&\Big[m_{_h}^2\Big]_{_{\phi\phi_{_{\tilde\eta}}}}=\frac{1}{2}g_{_B}g_{_{YB}}u\upsilon\left(\begin{array}{cc}
c_{\beta} c_{\beta^{\prime}},\;\;& -c_{\beta} s_{\beta^{\prime}}\\
c_{\beta^{\prime}} s_{\beta},\;\; &s_{\beta^{\prime}} s_{\beta}\end{array}\right)
\;,\nonumber\\
&&\Big[m_{_h}^2\Big]_{_{\phi_{_{\tilde\eta}}\phi_{_{\tilde\eta}}}}=\left(\begin{array}{cc}
g_{_B}^2 u^2 c_{\beta^{\prime}}^2+\rm{Re}(B_{\mu'}) t_{\beta^{\prime}},\;\;&
-g_{_B}^2u^2 s_{\beta^{\prime}}c_{\beta^{\prime}}-\rm{Re}(B_{\mu'})\\
-g_{_B}^2u^2 s_{\beta^{\prime}}c_{\beta^{\prime}}-\rm{Re}(B_{\mu'}),\;\; &
g_{_B}^2 u^2 s_{\beta^{\prime}}^2+\rm{Re}(B_{\mu'})/t_{\beta^\prime}\end{array}\right)\;,
\label{mh-tree-level1}
\end{eqnarray}
here abbreviations are $c_{\beta}=cos{\beta}$, $s_{\beta}=sin{\beta}$, $t_{\beta}=tan{\beta}$,
$c_{\beta^{\prime}}=cos{\beta^{\prime}}$, $s_{\beta^{\prime}}=sin{\beta^{\prime}}$ and
$t_{\beta^{\prime}}=tan{\beta^{\prime}}$.

\section{The radiative corrections $\Delta{\Pi^{(1)}_{hh}}$ and
$\Delta\Lambda_{_{abc}}(p_{_a}^2,\;p_{_b}^2,\;p_{_c}^2)$\label{app-B}}
\indent\indent
\begin{eqnarray}
&&\Delta{\Pi^{(1)}_{hh}}=\frac{-6{C^2_{hff}}}{{(4\pi)}^2}[(4m^2_f-m^2_h)B(m_f,m_f,m_h)-4m^2_f \ln{\frac{m^2_f}{Q^2}}]
\;,\nonumber\\
&&\Delta{\Pi^{(2)}_{hh}}=\frac{3{C^2_{h{s_i}{s_j}}}}{{(4\pi)}^2}[B(m_{s_i},m_{s_j},m_h)-B(m_{s_i},m_{s_j},0)]\;,
\label{self-corrections}
\end{eqnarray}
where $C_{_{XYZ}}$ is the coupling constant among the fields labeled by subscripts $X,\;Y,\;Z$, and
the function $B$ is defined as
\begin{eqnarray}
&&B(m_1,m_2,m_3)={\int^1_0}dx \ln\frac{(1-x){m^2_1}+x{m^2_2}+x(x-1){m^2_3}}{Q^2}\;.
\end{eqnarray}

\begin{eqnarray}
&&\Delta\Lambda_{_{h_{i}hh}}(p_{_a}^2,\;p_{_b}^2,\;p_{_c}^2)=
\sum\limits_{\alpha=1}^3\Delta\Lambda_{{h_{i}hh}}^{(\alpha)}
\end{eqnarray}
with
\begin{eqnarray}
&&\Delta\Lambda_{{h_{i}hh}}^{(1)}=-\frac{12m_f C_{{h_i}ff}C^2_{hff}}{(4\pi)^2}
\Big[B(m_f,m_f,m_{h_i})+2B(m_f,m_f,m_{h})
\nonumber\\
&&\hspace{1.5cm}
-3B(m_f,m_f,0)+(4m^2_f-\frac{m^2_{h_i}}{2}-m^2_h)B_1(m_f,m_f,m_f,m_h,m_{h_i})
\nonumber\\
&&\hspace{1.5cm}
-4m^2_fB_1(m_f,m_f,m_f,0,0)\Big]\;,
\label{trilinear-couplings 1}
\end{eqnarray}
with
\begin{eqnarray}
&&\Delta\Lambda_{{h_{i}hh}}^{(2)}=
-\frac{3 C_{{h_i}{S_i}{S_j}}C_{{h}{S_i}{S_j}}C_{{h}{S_j}{S_k}}}{(4\pi)^2}
\Big[B_1(m_{S_i},m_{S_j},m_{S_k},m_h,m_{h_i})
\nonumber\\
&&\hspace{1.5cm}
-B_1(m_{S_i},m_{S_j},m_{S_k},0,0)\Big]\;,
\label{trilinear-couplings 2}
\end{eqnarray}
and
\begin{eqnarray}
&&\Delta\Lambda_{{h_{i}hh}}^{(3)}=
\frac{3 C_{{h_i}{S_i}{S_j}}C_{hh{S_i}{S_j}}}{(4\pi)^2}\Big[B(m_{S_i},m_{S_j},m_{h_i})
-B(m_{S_i},m_{S_j},0)\Big]\;.
\end{eqnarray}
Where the function $B_1$ is defined as
\begin{eqnarray}
&&B_1(m_1,m_2,m_3,m_4,m_5)={\int^1_0}dx{\int^{1-x}_0}dy\Big[(1-x-y)m^2_1+xm^2_2+ym^2_3
\nonumber\\
&&\hspace{5.0cm}
+x(x-1)m^2_4+y(x+y-1)m^2_5\Big]^{-1}\;.
\label{trilinear-couplings 3}
\end{eqnarray}

\section{Some couplings\label{app-C}}
\indent\indent
The trilinear couplings between the lightest neutral Higgs and charged Higgs is written as
\begin{eqnarray}
&&C_{_{h_0H^+H^-}}=\frac{i}{4}\Big\{2(c_{_\beta}^2-s_{_\beta}^2)\Big(g_{_1}g_{_{BY}}+g_{_{YB}}g_{_B}\Big)
\Big[-\upsilon_{_{\bar\eta}}(Z_{_H})_{14}+\upsilon_{_\eta}(Z_{_H})_{13}\Big]
\nonumber\\
&&\hspace{2.2cm}
-(Z_{_H})_{11}\Big[(g_{_1}^2+g_{_{YB}}^2)(s_{_\beta}^2-c_{_\beta}^2)\upsilon_{_d}
+g_{_2}^2\upsilon_{_d}-2g_{_2}^2\upsilon_{_u}s_{_\beta}c_{_\beta}\Big]
\nonumber\\
&&\hspace{2.2cm}
+(Z_{_H})_{12}\Big[(g_{_1}^2+g_{_{YB}}^2)(s_{_\beta}^2-c_{_\beta}^2)\upsilon_{_u}
-g_{_2}^2\upsilon_{_u}+2g_{_2}^2\upsilon_{_d}s_{_\beta}c_{_\beta}\Big]\Big\}\;.
\label{RareB-9}
\end{eqnarray}
The couplings between the neutral gauge boson $Z^\prime$ and the SM fermions are
\begin{eqnarray}
&&g_{_L}^e=\frac{1}{2g_{_B}}\Big\{\Big[(g_{_1}+g_{_{BY}})s_{_W}-g_{_2}c_{_W}\Big]s_{_W}^\prime
+(g_{_{YB}}+g_{_B})c_{_W}^\prime\Big\}\;,
\nonumber\\
&&g_{_R}^e=-\frac{1}{2g_{_B}}\Big\{(2g_{_1}+g_{_{BY}})s_{_W}s_{_W}^\prime
+(2g_{_{YB}}+g_{_B})c_{_W}^\prime\Big\}\;,
\nonumber\\
&&g_{_L}^u=\frac{1}{6g_{_B}}\Big\{\Big[(g_{_1}+g_{_{BY}})s_{_W}-3g_{_2}c_{_W}\Big]s_{_W}^\prime
+(g_{_{YB}}+g_{_B})c_{_W}^\prime\Big\}\;,
\nonumber\\
&&g_{_R}^u=-\frac{1}{6g_{_B}}\Big\{(4g_{_1}+g_{_{BY}})s_{_W}s_{_W}^\prime
+(4g_{_{YB}}+g_{_B})c_{_W}^\prime\Big\}\;,
\nonumber\\
&&g_{_L}^d=-\frac{1}{6g_{_B}}\Big\{\Big[(g_{_1}+g_{_{BY}})s_{_W}+3g_{_2}c_{_W}\Big]s_{_W}^\prime
+(g_{_{YB}}+g_{_B})c_{_W}^\prime\Big\}\;,
\nonumber\\
&&g_{_R}^d=\frac{1}{6g_{_B}}\Big\{(2g_{_1}-g_{_{BY}})s_{_W}s_{_W}^\prime
+(2g_{_{YB}}-g_{_B})c_{_W}^\prime\Big\}\;.
\label{zPrime-3a}
\end{eqnarray}
The abbreviations
$c_{_W}=\cos\theta_{_W},\;s_{_W}=\sin\theta_{_W}$,
$c_{_W}^\prime=\cos\theta_{_W}^\prime,\;s_{_W}^\prime=\sin\theta_{_W}^\prime$
with $\theta_{_W}$ denoting the weinberg angle, and
\begin{eqnarray}
&&\sin^2\theta_{_W}^\prime=\frac{(g_{_1}^2+g_{_2}^2)g_{_{YB}}^2u^2}{16g_{_B}^4\upsilon^2}\;.
\label{zPrime-3b}
\end{eqnarray}

\section{Nonzero amplitudes from Fig.~\ref{Feynman diagrams}\label{app1}}
\indent\indent
Those nonzero amplitudes are summarized as
\begin{eqnarray}
&&M^{(1)}_{++}=\frac{{-i(\alpha_s{(\mu_{_R})})} C_{h_ihh}C_{h_iff} }{\pi[(\hat{s}-m^2_{h_i})+i m_{h_i}
\Gamma_{h_i}]}F^{(1/2)}_\triangle\;,
\nonumber\\
&&M^{(2+3+4)}_{++}=\frac{{i(\alpha_s{(\mu_{_R})})} C_{hff}^2 }{2 \pi \hat{s}}F^{(1/2)}_\Box\;,
\nonumber\\
&&M^{(2+3+4)}_{+-}=\frac{{-i(\alpha_s{(\mu_{_R})})} C_{hff}^2 }{2 \pi}G^{(1/2)}_\Box\;,
\nonumber\\
&&M^{(5+6)}_{++}=\frac{{i(\alpha_s{(\mu_{_R})})} C_{h_ihh}C_{h_{s_is_i}}}
{2 \pi[(\hat{s}-m^2_{h_i})+i m_{h_i}\Gamma_{h_i}]}F^{(0)}_\triangle\;,
\nonumber\\
&&M^{(7+8)}_{++}=\frac{{-i(\alpha_s{(\mu_{_R})})} C_{hh{s_is_i}}}{2 \pi} F^{(0)}_\triangle\;,
\nonumber\\
&&M^{(9)}_{++}=\frac{{-i(\alpha_s{(\mu_{_R})})} {C^2_{h s_i s_j}}}{\pi} F'^{(0)}_\triangle\;,
\nonumber\\
&&M^{(10+11+12)}_{++}=\frac{{i(\alpha_s{(\mu_{_R})})} {C^2_{h s_i s_j}}}{2 \pi \hat{s} } F^{(0)}_\Box\;,
\nonumber\\
&&M^{(10+11+12)}_{+-}=\frac{{i(\alpha_s{(\mu_{_R})})} {C^2_{h s_i s_j}}}{2 \pi } G^{(0)}_\Box\;.
\label{amplitudes}
\end{eqnarray}
Here $F_\triangle$ and $F'_\triangle$ are form factors associated with triangular diagrams.
$ F_\Box$ and $G_\Box$ are form factors of box diagrams which, respectively,
correspond to the same and opposite polarizations of the incoming gluons.
The concrete expressions of the form factors $(F_\triangle,F'_\triangle, F_\Box, G_\Box)$ are
\begin{eqnarray}
F^{(1/2)}_\triangle=2 m_f+(4m^2_f-\hat{s})m_f C_0(0,0,\hat{s},m^2_f,m^2_f,m^2_f)\;,
\end{eqnarray}
\begin{eqnarray}
&&F^{(1/2)}_\Box=-4 \hat{s}-8 m^2_f C_0(0,0,\hat{s},m^2_f,m^2_f,m^2_f) \hat{s}\nonumber\\
&&\qquad\quad-(8 m^2_f -2 m^2_h) [2(m^2_h- \hat{t}) C_0(0,m^2_h,\hat{t},m^2_f,m^2_f,m^2_f)\nonumber\\
&&\qquad\quad+2{(m^2_h- \hat{u})} C_0(0,m^2_h,\hat{u},m^2_f,m^2_f,m^2_f)\nonumber\\
&&\qquad\quad-(m^4_h - \hat{t} \hat{u}) D_0(0,m^2_h,0,m^2_h,\hat{t},\hat{u},m^2_f,m^2_f,m^2_f,m^2_f) ] \nonumber\\
&&\qquad\quad-2 m^2_f (8m^2_f-2m^2_h-\hat{s}) \hat{s}[D_0(0,m^2_h,0,m^2_h,\hat{t},\hat{u},m^2_f,m^2_f,m^2_f,m^2_f)\nonumber\\
&&\qquad\quad+D_0(0,0,m^2_h,m^2_h,\hat{s},\hat{t},m^2_f,m^2_f,m^2_f,m^2_f)\nonumber\\
&&\qquad\quad+D_0(0,0,m^2_h,m^2_h,\hat{s},\hat{u},m^2_f,m^2_f,m^2_f,m^2_f)\;,
\end{eqnarray}
\begin{eqnarray}
&&G^{(1/2)}_\Box=\frac{1}{(m^4_h-\hat{t} \hat{u})}\{[(8 m^2_f - \hat{t} - \hat{u})(2m^4_h -{\hat{t}}^2 - {\hat{u}}^2) C_0(m^2_h,m^2_h,\hat{s},m^2_f,m^2_f,m^2_f)\nonumber\\
&&\qquad\quad+(m^4_h-8m^2_f \hat{t} +{\hat{t}}^2)[2 (m^2_h-\hat{t})C_0(0,m^2_h,\hat{t},m^2_f,m^2_f,m^2_f)\nonumber\\
&&\qquad\quad-\hat{s}C_0(0,0,\hat{s},m^2_f,m^2_f,m^2_f)+\hat{s}\hat{t}D_0(0,0,m^2_h,m^2_h,\hat{s},\hat{t},m^2_f,m^2_f,m^2_f,m^2_f)]\nonumber\\
&&\qquad\quad+(m^4_h-8m^2_f \hat{u}+{\hat{u}}^2)[2(m^2_h-\hat{u})C_0(0,m^2_h,\hat{u},m^2_f,m^2_f,m^2_f)\nonumber\\
&&\qquad\quad-\hat{s}C_0(0,0,\hat{s},m^2_f,m^2_f,m^2_f)+\hat{s}\hat{u}D_0(0,0,m^2_h,m^2_h,\hat{s},\hat{u},m^2_f,m^2_f,m^2_f,m^2_f)]\nonumber\\
&&\qquad\quad+2m^2_f(m^4_h-\hat{t}\hat{u})(8m^2_f-\hat{t}-\hat{u})[D_0(0,m^2_h,0,m^2_h,\hat{t},\hat{u},m^2_f,m^2_f,m^2_f,m^2_f)\nonumber\\
&&\qquad\quad+D_0(0,0,m^2_h,m^2_h,\hat{s},\hat{t},m^2_f,m^2_f,m^2_f,m^2_f)\nonumber\\
&&\qquad\quad+D_0(0,0,m^2_h,m^2_h,\hat{s},\hat{u},m^2_f,m^2_f,m^2_f,m^2_f)]\}\;,
\end{eqnarray}
\begin{eqnarray}
F^{(0)}_\triangle=1+2m^2_s C_0(0,0,\hat{s},m^2_s,m^2_s,m^2_s)\;,
\end{eqnarray}
\begin{eqnarray}
F'^{(0)}_\triangle= C_0(m^2_h,m^2_h,\hat{s},m^2_{S_i},m^2_{S_j},m^2_{S_i})\;,
\end{eqnarray}
\begin{eqnarray}
&&F^{(0)}_\Box=(m^2_h-\hat{t}) C_0(m^2_h,0,\hat{t},m^2_{S_j},m^2_{S_i},m^2_{S_i})+(m^2_h-\hat{u})C_0(m^2_h,0,\hat{u},m^2_{S_i},m^2_{S_j},m^2_{S_j})\nonumber\\
&&\qquad\;+(m^2_h-\hat{t}) C_0(m^2_h,0,\hat{t},m^2_{S_i},m^2_{S_j},m^2_{S_j})+(m^2_h-\hat{u})C_0(m^2_h,0,\hat{u},m^2_{S_j},m^2_{S_i},m^2_{S_i})\nonumber\\
&&\qquad\;+2\hat{s}C_0(m^2_h,m^2_h,\hat{s},m^2_{S_i},m^2_{S_j},m^2_{S_i})\nonumber\\
&&\qquad\;+[(m^2_{S_j}-m^2_{S_i})\hat{s}-(m^4_h-\hat{t}\hat{u})]D_0(m^2_h,0,m^2_h,0,\hat{t},\hat{u},m^2_{S_j},m^2_{S_i},m^2_{S_i},m^2_{S_j})\nonumber\\
&&\qquad\;+2\hat{s}m^2_{S_i}[D_0(m^2_h,0,m^2_h,0,\hat{t},\hat{u},m^2_{S_j},m^2_{S_i},m^2_{S_i},m^2_{S_j})\nonumber\\
&&\qquad\;+D_0(m^2_h,m^2_h,0,0,\hat{s},\hat{t},m^2_{S_i},m^2_{S_j},m^2_{S_i},m^2_{S_i})\nonumber\\
&&\qquad\;+D_0(m^2_h,m^2_h,0,0,\hat{s},\hat{u},m^2_{S_i},m^2_{S_j},m^2_{S_i},m^2_{S_i})]\;,
\end{eqnarray}
\begin{eqnarray}
&&G^{(0)}_\Box=\frac{1}{m^4_h-\hat{t}\hat{u}}\{\hat{s}(2m^2_{S_i}-2m^2_{S_j}+\hat{t}+\hat{u})C_0(0,0,\hat{s},m^2_{S_i},m^2_{S_i},m^2_{S_i})\nonumber\\
&&\qquad\;-\hat{t}[2(m^2_h-\hat{t})C_0(m^2_h,0,\hat{t},m^2_{S_j},m^2_{S_i},m^2_{S_i})]-\hat{u}[2(m^2_h-\hat{u})C_0(m^2_h,0,\hat{u},m^2_{S_j},m^2_{S_i},m^2_{S_i})]\nonumber\\
&&\qquad\;-(m^2_h-\hat{t})(m^2_{S_i}-m^2_{S_j})[C_0(m^2_h,0,\hat{t},m^2_{S_i},m^2_{S_j},m^2_{S_j})+C_0(m^2_h,0,\hat{t},m^2_{S_j},m^2_{S_i},m^2_{S_i})]\nonumber\\
&&\qquad\;-(m^2_h-\hat{u})(m^2_{S_i}-m^2_{S_j})[C_0(m^2_h,0,\hat{u},m^2_{S_i},m^2_{S_j},m^2_{S_j})+C_0(m^2_h,0,\hat{u},m^2_{S_j},m^2_{S_i},m^2_{S_i})]\nonumber\\
&&\qquad\;+(2m^4_h-{\hat{t}}^2-{\hat{u}}^2)C_0(m^2_h,m^2_h,\hat{s},m^2_{S_i},m^2_{S_j},m^2_{S_i})\nonumber\\
&&\qquad\;+[-\hat{s}(m^2_{S_i}-m^2_{S_j})^2+(m^2_{S_i}+m^2_{S_j})(m^4_h-\hat{t}\hat{u})][D_0(m^2_h,0,m^2_h,0,\hat{t},\hat{u},m^2_{S_j},m^2_{S_i},m^2_{S_i},m^2_{S_j})\nonumber\\
&&\qquad\;+D_0(m^2_h,m^2_h,0,0,\hat{s},\hat{t},m^2_{S_i},m^2_{S_j},m^2_{S_i},m^2_{S_i})+D_0(m^2_h,m^2_h,0,0,\hat{s},\hat{u},m^2_{S_i},m^2_{S_j},m^2_{S_i},m^2_{S_i})]\nonumber\\
&&\qquad\;+[-\hat{s}{\hat{t}}^2-(m^2_{S_i}-m^2_{S_j})(2\hat{t}\hat{s}-(m^4_h-\hat{t}\hat{u}))]D_0(m^2_h,m^2_h,0,0,\hat{s},\hat{t},m^2_{S_i},m^2_{S_j},m^2_{S_i},m^2_{S_i})\nonumber\\
&&\qquad\;+[-\hat{s}{\hat{u}}^2-(m^2_{S_i}-m^2_{S_j})(2\hat{u}\hat{s}-(m^4_h-\hat{t}\hat{u}))]D_0(m^2_h,m^2_h,0,0,\hat{s},\hat{u},m^2_{S_i},m^2_{S_j},m^2_{S_i},m^2_{S_i})\}\;\nonumber\\.
\end{eqnarray}
$C_0$ and $D_0$ are scalar integrals, defined as
\begin{eqnarray}
&&C_0(p^2_1,p^2_2,(p_1+p_2)^2,m^2_1,m^2_2,m^2_3)\nonumber\\
&&=\frac{\mu^{4-D}}{i\pi^{D/2}\gamma_\tau}\int\frac{d^Dq}{[q^2-m^2_1][(q+p_1)^2-m^2_2][(q+p_1+p_2)^2-m^2_3]}\;,
\end{eqnarray}
\begin{eqnarray}
&&D_0(p^2_1,p^2_2,p^2_3,p^2_4,(p_1+p_2)^2,(p_3+p_4)^2,m^2_1,m^2_2,m^2_3,m^2_4)\nonumber\\
&&=\frac{\mu^{4-D}}{i\pi^{D/2}\gamma_\tau}\int\frac{d^Dq}{[q^2-m^2_1][(q+p_1)^2-m^2_2][(q+p_1+p_2)^2-m^2_3][(q+p_1+p_2+p_3)^2-m^2_4]}\;.\nonumber\\
\end{eqnarray}

Where
\begin{eqnarray}
\gamma_\tau=\frac{\Gamma^2(1-\varepsilon)\Gamma(1+\varepsilon)}{\Gamma(1-2\varepsilon)}\;,  D=4-2\varepsilon\;.
\end{eqnarray}
The  integrals are independent of $\mu$ in the limit $\varepsilon\rightarrow 0$.
Furthermore the functions $C_{ij}(\hat{s},z)$ ($ij=gg,\;gq,\;q\bar{q}$) are given as
\begin{eqnarray}
&&C_{gg}(\hat{s},z)=-zP_{gg}(z)\:\log\frac{\mu^2_{_F}}{\tau s}
\nonumber\\
&&\hspace{2.2cm}
+6[1+z^4+(1-z)^4]\:{(\frac{\log(1-z)}{1-z})_+}\:+d_{gg}(\hat{s},z)\:,
\nonumber\\
&&C_{gq}(\hat{s},z)=-\frac{z}{2}P_{gq}(z)\:\log\frac{\mu^2_{_F}}{\tau s(1-z)^2}+d_{gq}(\hat{s},z)\:,
\nonumber\\
&&C_{q\bar{q}}(\hat{s},z)=d_{q\bar{q}}(\hat{s},z)\:.
\label{QCD correction1}
\end{eqnarray}
Here $P_{ij}(z)$\:($i,j =g, q, \bar{q}$) are the specific Altarelli-Parisi
splitting functions ~\cite{AP splitting function}, and the concrete expressions
for the virtual corrections $C_{virt}(\hat{s})$ and the real corrections
$d_{ij}(\hat{s},z)$ can be found in Ref.~\cite{QCD-correction}, respectively.

\end{document}